# Confirmatory Factor Analysis

## A Case study

Vera Costa, Rui Sarmento
*FEUP, Portugal*

### ABSTRACT

*Confirmatory Factor Analysis (CFA) is a particular form of factor analysis, most commonly used in social research. In confirmatory factor analysis, the researcher first develops a hypothesis about what factors they believe are underlying the used measures and may impose constraints on the model based on these a priori hypotheses. For example, if two factors are accounting for the covariance in the measures, and these factors are unrelated to one another, we can create a model where the correlation between factor X and factor Y is set to zero. Measures could then be obtained to assess how well the fitted model captured the covariance between all the items or measures in the model. Thus, if the results of statistical tests of the model fit indicate a poor fit, the model will be rejected. If the fit is weak, it may be due to a variety of reasons. We propose to introduce state of the art techniques to do CFA in R language. Then, we propose to do some examples of CFA with R and some datasets, revealing several scenarios where CFA is relevant.*

Keywords: Confirmatory Factor Analysis, Statistics, R Language, Mathematics, Factor Analysis

### INTRODUCTION

CFA and Exploratory Factor Analysis (EFA) are interconnected statistical techniques. Sometimes, when some concepts relation is to be tested, the researcher uses CFA to test a hypothetical model of the system he/she is trying to propose. Thus, CFA helps in identifying the factor structure we believe the phenomena follows or is described by. In these situations, some variables may not measure what we thought they should. If the theoretical factor structure is not confirmed with CFA, EFA is the logical next step. EFA allows us to determine what the factor structure looks like according to how a particular sample of phenomena measurements behaves, for example, through the use of a survey to an audience. Therefore, EFA is essential to determine underlying constructs for a set of measured variables, and CFA might be used apriori for the test or simulation of the model we think best approaches a specific concept or phenomena and then tests the hypothesis statistically.

Another possible approach, using both CFA and EFA, is to leverage the potential confirmation of the CFA after using EFA. Thus, by identifying factors that explain the majority of variance with EFA, we can confirm the model with the statistical tests available for CFA.

Some care should be taken when using CFA; the results change considerably when the hypothesis being tested is changed, even when just a little bit. Additionally, when some of the following list of requirements are discured, results might diverge:

- multivariate normality;
- good parameter identification;
- processing of outliers;
- processing of missing data.

CFA vs. EFA

- CFA and EFA are linear statistical models;
- CFA and EFA assume a normal distribution;
- CFA and EFA incorporate measured variables;
- CFA requires specification of a model;
- CFA requires specification of the number of factors (theoretical, empirical or after EFA);
- CFA requires specification of which variables load on each factor (theoretical, empirical or after EFA),
- CFA requires specification of error explicitly.

## BACKGROUND

### Exploratory factor analysis

Exploratory Factor Analysis (EFA) is a statistical method used to describe variability among observed, correlated variables. The goal of performing exploratory factor analysis is to search for some unobserved variables called factors (Rui Sarmento & Costa, 2017). EFA analysis might lead to the conclusion that a reduced number of unobserved latent variables are reflected in the variations of a high number of observed variables. Observed variables are modeled as linear combinations of the possible factors, added the error quantification of this approximation.

EFA should start with the analysis of the correlation matrix. Depending on the variable type, different methods to obtain this matrix could be used: Pearson (for quantitative variables), Spearman (for ordinal variables) and Cramer's V (for nominal variables). Based on the correlation matrix, the researcher frequently discusses the existence or non-existence of at least two factors.

*Sampling Adequacy*

**Bartlett Sphericity test**

Bartlett's test is used to test if several samples have equal variances. If so, this is called homogeneity of variances. In some statistical tests, as is the case of the analysis of variance, it is assumed that variances are equal between groups or samples. Bartlett test can be used to analyze that assumption. Thus, the hypotheses of this test are (Rui Sarmento & Costa, 2017):

$H_0$: *the matrix of population correlations is equal to the identity matrix*
$H_1$: *the matrix of population correlations is different from the identity matrix*

**KMO Measure**

Kaiser-Meyer-Olkin (KMO) measure tests the sampling adequacy for each variable in the model and the complete model. This statistic is a measure of the proportion of variance among variables that might be common variance. KMO checks if it is possible to factorize the primary variables efficiently.
Thus, for reference, Kaiser suggested the following classification of the results (Rui Sarmento & Costa, 2017):

- 0 to 0.49 unacceptable
- 0.50 to 0.59 miserable
- 0.60 to 0.69 mediocre
- 0.70 to 0.79 middling
- 0.80 to 0.89 meritorious
- 0.90 to 1.00 marvelous.

**Number of factors to be retained**

To perform EFA, the researcher should know how many factors should be maintained. Several methods are available to decide it:

- Kaiser criterion: according to this rule, only factors with eigenvalues higher than one are retained for interpretation;
- Scree plot: involves the visual exploration of a graphical representation of the eigenvalues. The point where the last significant drop or break takes place is used to define the number of factors;
- Variance explained criteria: this method consists in to retain the number of factors that account for a certain percent of the extracted variance. Depending on the research area, different values of the percent of the extracted variance could be defined.

**Methods of communalities' estimates**

Diverse methods could be used to estimate the communalities. Based on this estimation, it is possible to obtain a matrix of factor weights. In this sense, some methods commonly used are (R. Sarmento & Costa, 2017):

- Principal Component Method
- Principal Axis Method of Factor Extraction
- Maximum Likelihood Method (ML)

**Factor Rotation**

EFA solution is not always interpretable. The factor weights of the variables in common factors can be such that it is not possible to assign a meaning to extracted empirical factors. Thus, there are several methods to make the rotation of the factorial axes, such as:

- Varimax method
- Quartimax method
- Oblimin method

## Confirmatory factor analysis

### *Reliability and Validity*

After performing the EFA, it is necessary to confirm the obtained results, i.e., perform the Confirmatory Factor Analysis (CFA). More specifically CFA is a technique that "seeks to confirm if the number of factors (or constructs) and the loadings of observed (indicator) variables on them conform to what is expected on the basis of theory" (Malhotra, Hall, Shaw, & Oppenheim, 2007). Thus, to reach the confirmation and to accurately perceive the representation of the constructs by the observed variables, it is necessary to evaluate the reliability and validity of the scale (Joseph F. Hair, Black, Babin, Anderson, & Tatham, 2009). The most commonly used method is Cronbach's Alpha, which measures reliability and internal consistency. A commonly accepted rule for describing internal consistency using Cronbach's alpha is (Rui Sarmento & Costa, 2017):

- 0 to 0.49 unacceptable
- 0.50 to 0.59 poor
- 0.60 to 0.69 questionable
- 0.70 to 0.79 acceptable
- 0.80 to 0.89 good
- from 0.9 to 1 excellent

However, other alternatives have been adopted, such as Construct Reliability (CR). Hair (2009) defines CR as the "measure of reliability and internal consistency of measured variables that represent a latent construct." The CR should be measured before the validity of the construct be evaluated.

Regarding the validity, it is intended to estimate if the scale measures or operationalizes the construct that the researcher wishes to evaluate in reality. There are several types of validity methods which varies according to the research objectives. The main techniques are convergent validity and discriminant validity.

Convergent validity is based on the study of expected and plausible relationships with other measures related to two types of variables: i) relationship established with variables measured by different instruments that intentionally measure the same construct; ii) relationships with instruments that measure other aspects with which a positive or negative relationship is expected to exist (Silva, Macêdo, & Silva, 2013). This type of validity can be explained by Construct Reliability (CR) and Average Variance Extracted (AVE), a measure of consistency that reveals the mean percentage of explained variance between the items of a construct. According to Hair (2009), the convergent validity is observed when CR is higher than the AVE, and the AVE is higher than 0.5.

Discriminant validity, also mentioned as divergent validity, consists of the degree "to which a measure does not correlate with other measures from which it is assumed to diverge" (Sánchez, 1999). Additionally, careful planning of the validation process should be carried out during the preparation of the instrument to collect the necessary data simultaneously. Thus, in the validation process, it is required to specify the predicted hypotheses among the involved variables indicating: a) the expected meaning of the relationship, whether positive, negative, or lacking relationship and ii) the expected relative magnitude of the association, where it can be proved that there are more significant and more transparent relationships (Silva et al., 2013). This type of validity can be seen in the diagonal matrix that shows the square root of the AVE - there is a discriminant validity when this is higher to the correlations of the construct under analysis.

After checking and finalizing these points, it will be possible to advance to the structural model and, consequently, to the results of the whole analysis of the Structural Equations Model.

### *Model specification*

Two types of models should be explored in CFA: the **measurement model** and the **structural model**. The **measurement model** concerns the relations between measures of constructs, indicators, and the constructs they were designed to measure (i.e., factors). By examining three critical sets of results – parameter estimates, fit indices, and, potentially, modification indices – researchers formally test measurement hypotheses, and they can modify hypotheses to be more consistent with the actual structure of participants' responses to the scale. Thus, it is crucial starting by defining the measurement model.

To do this, researchers should start to specify at least three essential facets of a measurement model. First, they should determine the number of factors or latent variables (represented by ovals) hypothesized to underlie the scale's items (represented by rectangles). Second, they should specify the items linked to each factor, with at least one item related to each factor, and with each item linked to only one latent variable. Third, if a hypothesized model includes multiple factors, researchers should specify possible associations between factors (Furr, 2011). In this sense, some authors (Leach et al., 2008) present an example of a measurement model (Figure 1). This specification implies that two factors – Self-definition and Self-investment – are hypothesized to be correlated.

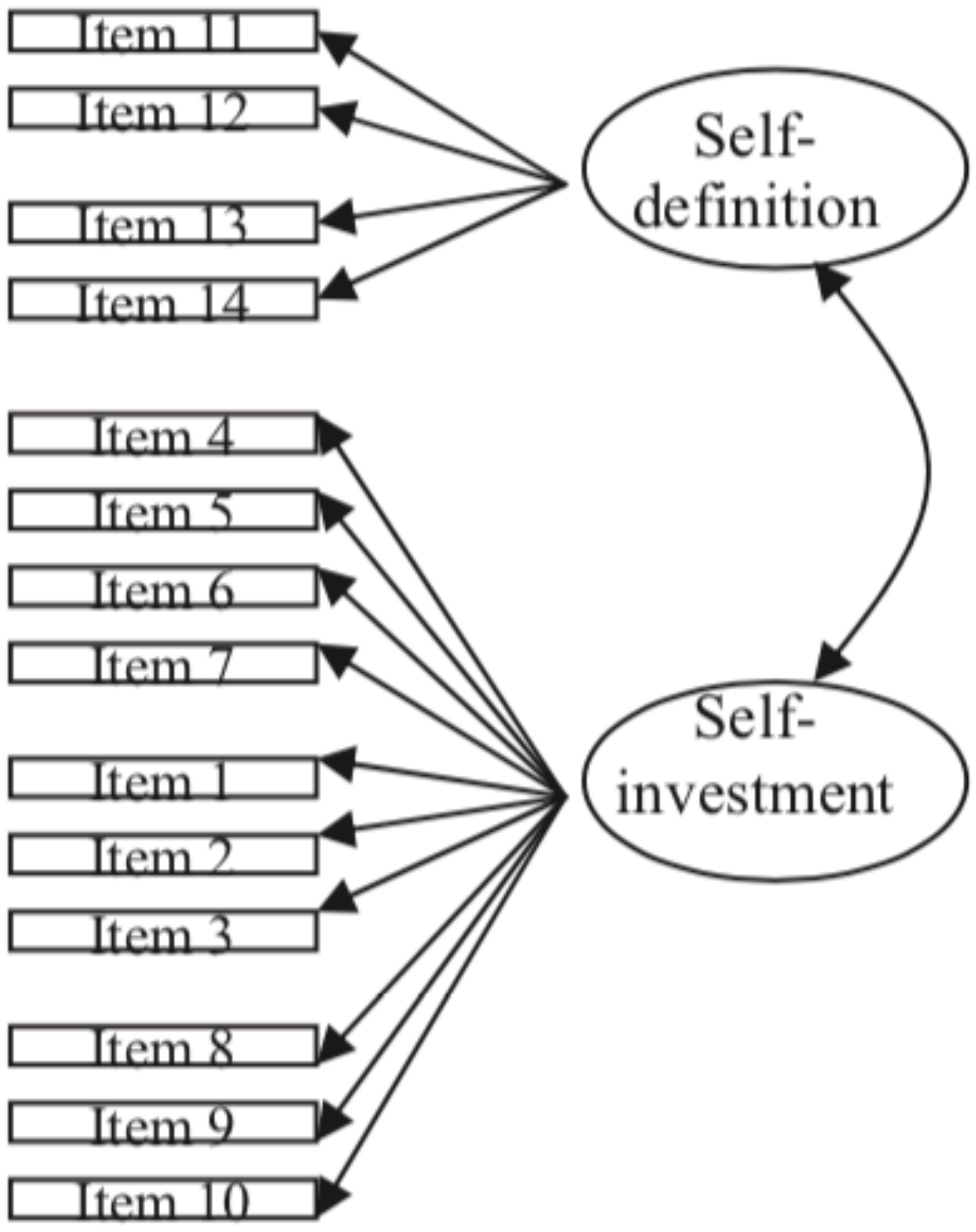

*Figure 1. Example of a measurement model (Leach et al., 2008)*

Another example is given by (Lewis, 2017). The author assumes that a researcher would like to test the construct validity of a 10-item instrument that measures social interest and lifestyle. His previous research suggests that two latent variables would be reasonable to include in the model: one latent variable for social interest and one latent variable for lifestyle. These latent variables are proposed to be defined by the observed (or indicator) variables, which in this case are specific items in the assessment. He assumes items 1, 3, 7, 9, and 10 are specified to load on the social interest latent variable, and items 2, 4, 5, 6, and 8 are defined to load on the lifestyle latent variable. It is also presented the measurement errors for each variable. Thus, his proposed model is shown in Figure 2.

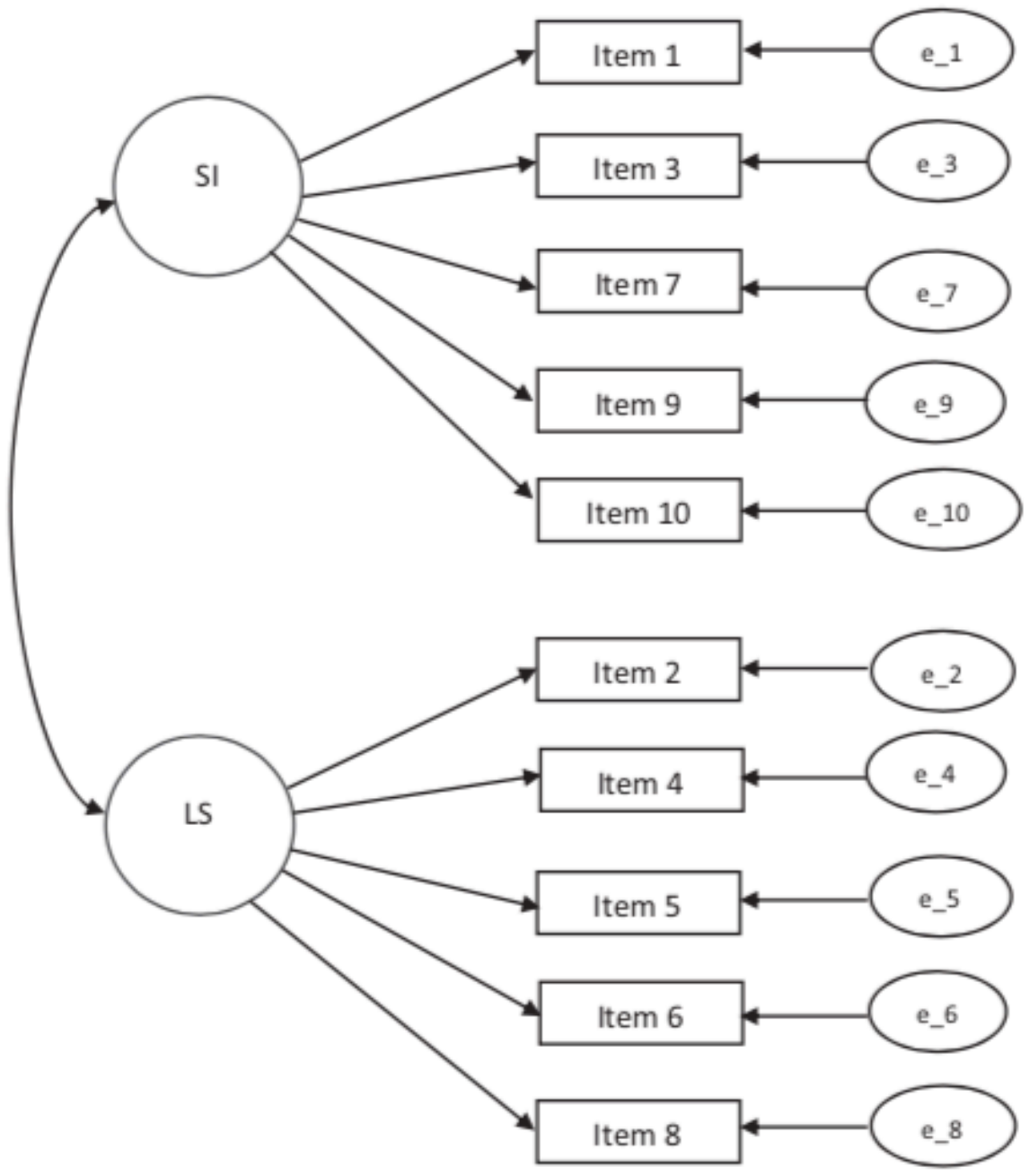

*Figure 2. Example of a measurement model with two factors (Lewis, 2017)*

After the elaboration of the measurement model, it is necessary to proceed with its specification, i.e., the **structural model**. The specification of the model is its formal "drawing," which reflects, a priori, the assumptions about the measurement model. To this end, some rules must be followed. Latent common factors cause overt variables. The behavior of the manifested variables results from the manifestation of latent factors. Latent specific factors explain the variance of the displayed variables that are not defined by common latent factors (e.g., measurement errors). Measurement errors are generally independent (but may be correlated indicating a source of mutual variation of items not explained by common factors present in the model).

To synthesize, Figure 3 shows the measurement model and structural model, as well as its relationship.

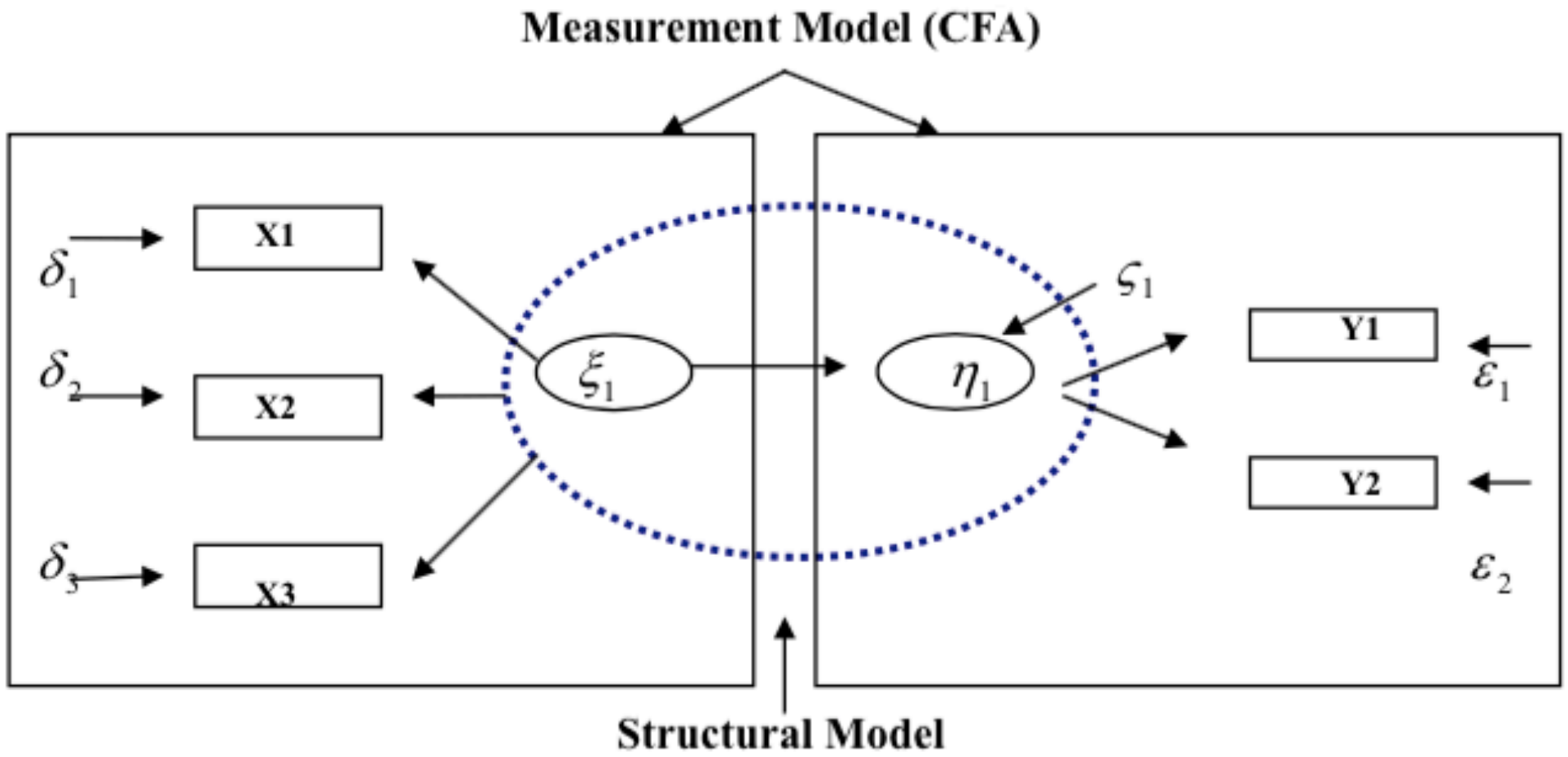

*Figure 3. Relation between measurement and structural models* (Gutierrez, 2005)

*Quality of Model Adjustment*

**Chi-squared test ($X^2$)**

Chi-squared test ($X^2$) indicates the difference between observed and expected covariance matrices. Hypotheses of this model are (Gunzler & Morris, 2016):

$H_0$: *the proposed model and the data structure are similar (no differences)*
$H_1$: *there is a difference between the proposed model and the data structure*

Values nearer to zero indicate a better fit, i.e., a smaller difference between expected and observed covariance matrices. On the opposite, a sizeable chi-squared test with a corresponding small $p - value$ indicates that the model does not fit the data (Suhr, 2006).

**Normed Fit Index (NFI)**

Normed fit index (NFI) is also called Bentler-Bonett Normed Fit Index. It analyzes the discrepancy between the chi-squared value of the proposed model and the chi-squared value of the null model. NFI tends to be negatively biased. It is considered very good if it is equal to or greater than 0.95, good between 0.9 and 0.95, suffering between 0.8 and 0.9 and bad if it is less than 0.8 (Portela, 2012).

$$NFI = 1 - \frac{X^2\ (proposed\ model)}{X^2\ (null\ model)}$$

**Comparative Fit Index (CFI)**

Comparative fit index (CFI) analyzes the model fit by examining the discrepancy between the data and the proposed model while adjusting for the issues of sample size intrinsic in the chi-squared test, and the normed fit index. It is considered very good if it is equal to or greater than 0.95, good between 0.9 and 0.95, suffering between 0.8 and 0.9 and bad if it is less than 0.8 (Portela, 2012).

$$CFI = 1 - \frac{\max\ [X^2_{proposed\ model} - df_{proposed\ model}, 0]}{\max\ [X^2_{null\ model} - df_{null\ model}, 0]}$$

**Relative Fit Index (RFI)**

Relative fit indices (RFI) is also called “incremental fit indices” or “comparative fit indices.” It compares the chi-square for the proposed model to a null model. This null model almost always contains a model in which all of the variables are uncorrelated, and as a result, has a very large chi-square (indicating poor fit). It is considered very good if the nearest is 1 and bad if it is less than 0,9 (Portela, 2012).

$$RFI = 1 - \frac{X^2_{proposed\ model}/df_{proposed\ model}}{X^2_{null\ model}/df_{null\ model}}$$

**Tucker-Lewis Index (TLI)**

Tucker-Lewis index (TLI) is also known as a non-normed fit index (NNFI). It is a combination of a measure of parsimony with a comparative index between the proposed model and the null model. It is considered very good if it is equal to or greater than 0.95, good between 0.9 and 0.95, suffering between 0.8 and 0.9 and bad if it is less than 0.8 (Portela, 2012).

$$TLI = \frac{\frac{X^2_{null\ model}}{df_{null\ model}} - \frac{X^2_{proposed\ model}}{df_{proposed\ model}}}{\frac{X^2_{null\ model}}{df_{null\ model}} - 1}$$

**Root Mean Square Error of Approximation (RMSEA)**

Root Mean Square Error of Approximation (RMSEA) is a measure that attempts to correct the tendency of chi-square statistics to reject models with large samples. It avoids issues of sample size by analyzing the discrepancy between the proposed model, with optimally chosen parameter estimates, and the population covariance matrix. RMSEA is considered very good if it is equal to or less than 0,05, good between 0,05 and 0,08, mediocre between 0,08 and 0,10 and unacceptable if it is higher than 0,10 (Portela, 2012).

$$RMSEA = \sqrt{max\left(\frac{X^2_{proposed\ model} - df_{proposed\ model}}{df_{proposed\ model} \times (N-1)}, 0\right)}$$

Where N is the sample size and $df$ the degrees of freedom. Additionally, RMSEA provides a one-sided test with the following hypotheses (Maccallum, Browne, & Sugawara, 1996):

$H_0$: *the RMSEA equals 0.05 (what is called a close-fitting model)*
$H_1$: *the RMSEA is higher than 0.05*

Thus,

- if $p - value \geq 0.05$ (i.e., not statistically significant), the fit of the model is "close."
- if $p - value \leq 0.05$, the fit of the model is worse than close fitting (i.e., the RMSEA is higher than 0.05).

*Table 1. Summary of reference values for adjustment indices*

| | Very Good | Good | Suffering | Bad |
|---|---|---|---|---|
| $X^2/df$ | $\leq 1$ | $]1,2]$ | $]2,5]$ | $> 5$ |
| NFI | $\geq 0.95$ | $[0.9; 0.95[$ | $[0.8; 0.9[$ | $< 0.8$ |
| CFI | $\geq 0.95$ | $[0.9; 0.95[$ | $[0.8; 0.9[$ | $< 0.8$ |
| RFI | the better the closer to 1 | | | |
| TLI | $\geq 0.95$ | $[0.9; 0.95[$ | $[0.8; 0.9[$ | $< 0.8$ |
| RMSEA ($p - value \geq 0.05$) | $\leq 0.05$ | $]0.05, 0.08]$ | $]0.08,0.10]$ | $> 0.10$ |

## MAIN FOCUS OF THE CHAPTER

### Data

An example of confirmatory factor analysis will be given in this chapter. Data from a survey presented to 204 researchers will be considered. The goal of the surveys was to understand the researcher's behavior. The first six rows of data are given by:

```
head(data.df)
```

Corresponding output is represented in Table 4 in Appendix.
Description of variables should be performed in this type of analysis, which allows us discovering irregularities (for example, missing values or outliers). Thus, descriptive analysis of Q1 to Q21 variables is given by:

```
# Identification of the variables used in factor analysis
survey<-data.df[, paste("Q", 1:21, sep="")]

# Descriptive analysis for each variable
summary(survey)
```

Table 5 (in the appendix) shows the output of descriptive analysis. It is possible to observe the non-existence of missing values since no count of "NAs" is presented. As answers are represented on a Likert scale, outliers should not exist, except when there are data errors. In this case, errors should be corrected, or the particular researcher's row data should be eliminated. Additionally, in case of missing values, the replacing by the mean or median of the corresponding variable could be made. However, some authors argue that the replacement can only occur when they are, in maximum, 20% of the total sample. As it can be seen, all variables vary between 1 and 7.

To finish the understanding of variables Q1 to Q21 the correlation matrix should be analyzed. Since Q1 to Q21 variables are ordinal, Spearman's correlation should be used (Rui Sarmento & Costa, 2017), and it is obtained using the code:

```
### Correlation between variables Q1 to Q10
correlation <- cor(survey, method="spearman")
correlation
```

The output (Table 6) show the correlations' matrix for Q1 to Q21 variables. The correlation values varies from 0.090 to 0.868 (excluding diagonal). This disparity of values shows that at least two factors should be used for the reduction of Q1 to Q21 variables. Thus, two or more factors will be considered, and the exploratory factor analysis makes sense with this database.

### Methodology

The present study has two main objectives: to uncover the underlying structure of our set of variables, and test whether measures of a construct are consistent with proposed researcher understanding of the nature of each factor, i.e., to test whether the data fits a proposed measurement model. The R code used to make all analysis is only an example. Probably, other packages or other expressions could be used.

To perform EFA the following steps will be used:

- determine number of factors to retain (with, for example, scree test or cumulative proportion);
- rotation – a transformation;
- interpret solution;
- calculate factor scores.

To perform EFA the following steps will be used (Suhr, 2006):

- review the relevant theory and research literature to support model specification;
- specify a model (e.g., diagram, equations);
- determine model identification;
- collect data;
- conduct preliminary descriptive statistical analysis;
- estimate parameters in the model;
- assess model fit;
- present and interpret the results.

## Results

### *Exploratory factor analysis*

Exploratory factor analysis starts with the application of two methods: Bartlett sphericity test and the KMO Measure (Rui Sarmento & Costa, 2017). Both methods could be used in R using PSYCH package.

The programming code for the Bartlett test and corresponding output are:

*Computer Code*

```
### Bartlett Sphericity test
library (psych)
cortest.bartlett(correlation, n=nrow(data.df))
```

*Output*

```
$chisq
[1] 3326.251

$p.value
[1] 0

$df
[1] 210
```

Previous output shows the rejection of the null hypothesis ($p < 0.05$) of the Bartlett test, i.e., the matrix of population correlations is different from the identity matrix. Thus, we can conclude that factor analysis is appropriate to our data.

KMO measure is calculated similarly to Bartlett test:

| | |
|---|---|
| *Computer Code* | `### KMO Measure`<br>`library (psych)`<br>`KMO(correlation)` |
| *Output* | `Kaiser-Meyer-Olkin factor adequacy`<br>`Call: KMO(r = correlation)`<br>`Overall MSA =  0.91`<br>`MSA for each item =`<br>`  Q1   Q2   Q3   Q4   Q5   Q6   Q7   Q8   Q9  Q10  Q11  Q12  Q13  Q14`<br>`0.92 0.92 0.95 0.96 0.91 0.91 0.95 0.94 0.96 0.87 0.77 0.88 0.81 0.79`<br>` Q15  Q16  Q17  Q18  Q19  Q20  Q21`<br>`0.96 0.89 0.91 0.92 0.91 0.87 0.90` |

The output shows a marvelous KMO value (0.91), which means that EFA could be performed. Additionally, all variables present values of KMO higher than 0.5. Thus, all variables could be considered for EFA. If some variable presents KMO smaller than 5, it should be removed or discussed in detail by the analyst.

Seeing that EFA is appropriate for the presented data, the analysis of factors to be retained is explored. As described in (R. Sarmento & Costa, 2017), several methods could be used to define the number of factors. Following is the first method: Kaiser criteria.

```
### Kaiser criterion
library (psych)
eigen(correlation)
```

Table 7 presents the outputs divided into "values" and "vectors" which corresponds to eigenvalues and eigenvectors, respectively. Thus, four eigenvalues higher than one are observed. This means that accordingly this method, four factors should be retained.
The second method to be used is the scree plot:

| | |
|---|---|
| *Computer Code* | `### Scree plot criterion`<br>`library(nFactors)`<br>`scree(correlation, hline=-1) # hline=-1 draw a horizontal line at -1` |
| *Output* | 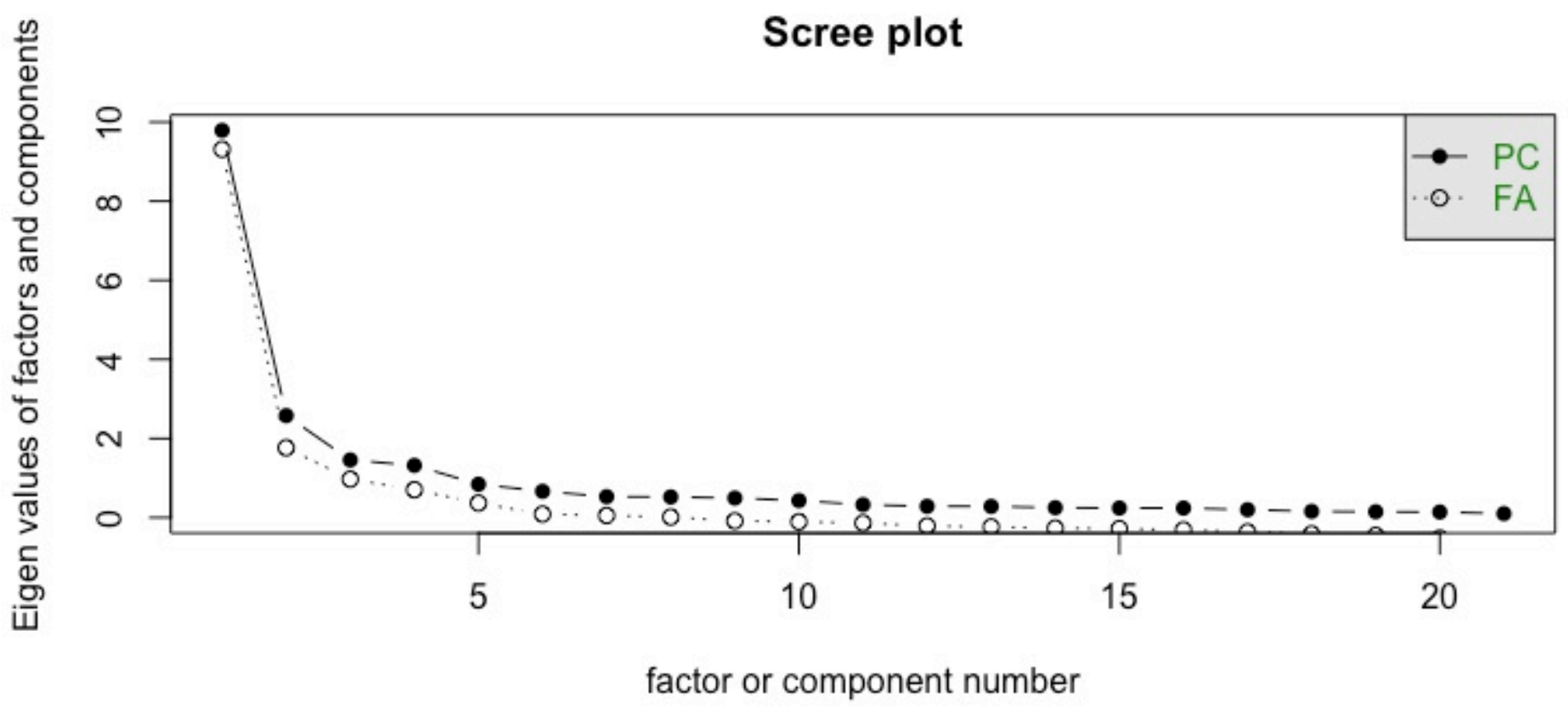 |

Observing the previous scree plot and curves make an "elbow" toward a less steep decline in value 3. Thus, this method suggests retaining three factors. Since the scree plot is a visual method, some doubts could arise.
Another method can be used to help in the decision of the number of factors to be retained: variance explained criteria:

| | |
|---|---|
| *Computer Code* | <pre>### Explained variance for each component<br>pc <- prcomp(survey,scale.=F)<br>summary(pc)</pre> |
| *Output* | <pre>Importance of components:<br>                          PC1     PC2     PC3     PC4     PC5     PC6<br>Standard deviation     5.2041 2.19297 1.93647 1.69202 1.44830 1.21755<br>Proportion of Variance 0.5257 0.09336 0.07279 0.05558 0.04072 0.02878<br>Cumulative Proportion  0.5257 0.61909 0.69189 0.74746 0.78818 0.81696<br>                           PC7     PC8     PC9    PC10    PC11<br>Standard deviation     1.14315 1.03692 1.03248 0.91380 0.87686<br>Proportion of Variance 0.02537 0.02087 0.02069 0.01621 0.01493<br>Cumulative Proportion  0.84232 0.86320 0.88389 0.90010 0.91503<br>                          PC12    PC13    PC14    PC15    PC16<br>Standard deviation     0.87483 0.82893 0.75825 0.68115 0.66079<br>Proportion of Variance 0.01486 0.01334 0.01116 0.00901 0.00848<br>Cumulative Proportion  0.92988 0.94322 0.95438 0.96339 0.97187<br>                          PC17    PC18    PC19   PC20    PC21<br>Standard deviation     0.62693 0.59443 0.54742 0.5125 0.37496<br>Proportion of Variance 0.00763 0.00686 0.00582 0.0051 0.00273<br>Cumulative Proportion  0.97950 0.98635 0.99217 0.9973 1.00000</pre> |

Previous output presents the importance of each component for the 21 variables in the study, i.e., the variance explained depending on the number of factors to be considered.
The minimum acceptable value of variance explained, according to several authors, is 75% (R. Sarmento & Costa, 2017). Thus, with four factors, approximately 75% of the variance is explained. However, the researcher must have critical thought and check if the number of suggested factors makes sense in the scope of the problem that is being analyzed. In our case study, four factors will be considered.

To analyze each factor, as well as the variables belonging to each factor, several methods could be used. The Principal Component method is used in this study:

```
### Principal Component method
library (psych)
principal(correlation,nfactors=4, rotate="none")
```

The output corresponding to the previous R code is presented in Table 8 (in the appendix). First four columns of the standardized loadings (pattern matrix) based upon correlation matrix, in the output, give us the variable's weight in each defined component. These weights allow us defining the variables belonging to each component, i.e., the variable should belong to the factor where it has the highest weight. Thus, following this rule, no variable would remain in factor 3. Also, the Q20 variable raises doubts because the weight in factors 1 and 4 is 0.61. To eliminate these doubts, the results should be analyzed after a factor rotation. In this study, a varimax rotation will be considered:

```
### Principal Component method with varimax rotation
library (psych)
principal(correlation,nfactors=4, rotate="varimax")
```

In this case (Table 9 in the appendix), no doubts remain. Factor 1 contains nine variables: Q1 to Q9. In the second factor, five variables: Q10 to Q14. The third factor has four variables: Q15 to Q18. Finally, fourth has three variables: Q19 to Q21.
The h2 column of the output is also essential. It represents the values of communalities which must be higher than 0.3. In case of lower values, the variable should be excluded from the model and the analyses reperformed.

Additionally, since the RMSR value is lower than 0.1 ($RMSR = 0.05$), the retained factors are appropriate to describe the correlation structure.

### *Confirmatory factor analysis*

As discussed above (background section), to begin the confirmatory factor analysis, the researcher should have a model in mind. The idea of this model can be drawn from the literature review or the exploratory factorial analysis previously done. Thus, in the present case study, the model resulted from the EFA will be used.

**Reliability of the model**

To perform CFA, the reliability and validity of the model should be analyzed. Several measures could be used. Internal consistency of Cronbach's alpha (α) is one of these measures. The coefficient (α) varies between 0 and 1, and measures the degree to which the items in an array of data are correlated.

Thus, the reliability analysis for the first factor is given by:

```
# PC1 (Q1, ..., Q9)
library (psych)
alpha(survey[c(paste("Q",1:9,sep=""))])
```

Table 10 (in the appendix) show the results obtained for factor 1. As it is possible to verify, the value of alpha (raw_alpha) is equal to 0.94, which means that this is an "excellent" value, according to some authors. The values of the "reliability if an item is dropped" show a lower or equal alpha value for all variables of this factor. This means that all of them contribute positively to the internal consistency of the factor. Hence, we can conclude that the first factor is well defined.

Regarding the second factor, the internal consistency is given by:

```
# PC2 (Q10...Q14)
library (psych)
alpha(survey[c(paste("Q",10:14,sep=""))])
```

In the second factor (Table 11 in the appendix), $\alpha = 0.88$, i.e., a "good" value. Additionally, if some item/variable of this factor is dropped, the value of alpha decreases. Thus, it can be concluded that the second factor is well defined.

The internal consistency of the third factor is given by:

```
# PC2 (Q15...Q18)
library (psych)
alpha(survey[c(paste("Q",15:18,sep=""))])
```

We can observe an alpha value is equal to 0.91 (Table 12 in the appendix), which means that this is an "excellent" value. Additionally, if some item/variable of this factor is dropped, the value of alpha decreases, which means that all variables contribute positively to the factor. Thus, the third factor is also well defined.

Finally, the internal consistency of the fourth factor is given by:

```
# PC2 (Q19...Q21)
library (psych)
alpha(survey[c(paste("Q",19:21,sep=""))])
```

This factor (Table 13, in the appendix) presents a "good" value of the internal consistency ($\alpha = 0.83$). Similar to previous cases, if some item/variable is dropped, the alpha values decreases. Again, it can be concluded that the third factor is also well defined.
To point out that if some alpha value increases (considering "if an item is dropped"), this means the internal consistency of the factor is better without the dropped variable. In that case, the variable is harming the factor and should be removed to the analysis.

To summarize, the following table (Table 2) presents the values of reliability for each factor, as well as for the factor if the variable is dropped.

*Table 2. Reliability and importance of each factor*

| | **Number of Items** | **Reliability of the factor ($\alpha$)** | **Variables** | **Reliability if an item is dropped ($\alpha$)** |
|---|---|---|---|---|
| **Factor 1** | 9 | 0.94 | Q1 | 0.93 |
| | | | Q2 | 0.93 |
| | | | Q3 | 0.94 |
| | | | Q4 | 0.94 |
| | | | Q5 | 0.94 |
| | | | Q6 | 0.94 |
| | | | Q7 | 0.94 |
| | | | Q8 | 0.94 |
| | | | Q9 | 0.94 |
| **Factor 2** | 5 | 0.88 | Q10 | 0.87 |
| | | | Q11 | 0.86 |
| | | | Q12 | 0.86 |
| | | | Q13 | 0.85 |
| | | | Q14 | 0.86 |
| **Factor 3** | 4 | 0.91 | Q15 | 0.89 |
| | | | Q16 | 0.86 |
| | | | Q17 | 0.90 |
| | | | Q18 | 0.89 |
| **Factor 4** | 3 | 0.83 | Q19 | 0.82 |
| | | | Q20 | 0.70 |
| | | | Q21 | 0.76 |

**Convergent validity**

As explained in previous section, convergent validity can be explained by Construct Reliability (CR) and Average Variance Extracted (AVE):

<table>
<tr><td>Computer Code</td><td>

```
pca.varimax <- principal(correlation,nfactors=4, rotate="varimax")

pca.coefs <- matrix(as.numeric(pca.varimax$loadings[1:length(pca.varimax
$loadings)]),nrow=21,ncol=4,byrow=FALSE)

nfactor <- apply(pca.coefs,1,FUN=function(x){
  which.max(x)
  })
crave <- data.frame(matrix(ncol=2,nrow=0),stringsAsFactors=FALSE)

sapply(unique(nfactor), function(x){
    coeff  <- pca.coefs[which(x==nfactor),x]
    coeff2 <- pca.coefs[which(x==nfactor),x]^2
    e2 <- 1-coeff2
    cr <- sum(coeff)^2 / (sum(coeff)^2+sum(e2))
    ave <- sum(coeff2) / (sum(coeff2)+sum(e2))
    crave <<- rbind(crave,c(cr,ave))
} )
names(crave) <- c("CR","AVE")
crave
```

</td></tr>
<tr><td>Output</td><td>

```
         CR       AVE
1 0.9218027 0.5688222
2 0.8873831 0.6122406
3 0.8726442 0.6318886
4 0.8292765 0.6187450
```

</td></tr>
</table>

The presented output gives CR and AVE values. As mentioned in background section, convergent validity is observed when CR is higher than the AVE, and the AVE is higher than 0.5. These conditions are verified in the present case study and, consequently, convergent validity is verified.

**Discriminant (divergent) validity**

This type of validity can be seen in the diagonal matrix that shows the correlations between factors and square root of the AVE in the diagonal:

<table>
<tr><td>Computer Code</td><td>

```
library(psych)
library(GPArotation)
m.cor.ave <- fa(correlation, nfactors=4)$score.cor

for (i in 1:4){
+   m.cor.ave[i,i] <- sqrt(crave[i,"AVE"])
+ }
m.cor.ave
```

</td></tr>
<tr><td>Output</td><td>

```
           [,1]       [,2]       [,3]       [,4]
[1,] 0.75420303 0.38879189 0.66329573 0.59926812
[2,] 0.38879189 0.78245805 0.29289469 0.32172374
[3,] 0.66329573 0.29289469 0.79491418 0.52559418
[4,] 0.59926812 0.32172374 0.52559418 0.78660345
```

</td></tr>
</table>

This output present the correlation matrix between factors, with the diagonal with square root of the AVE. Discriminant validity is verified when the diagonal is higher to the correlations of the construct under analysis. This condition is verified and, thus, we can conclude that the model presents discriminant validity.

**Measurement model specification**

After verifying the convergent and discriminant validity of the constructs, the measurement model specification should be studied.

The following R code asks some quality measures for the measurement model. Results are presented in Table 14 in appendix.

```
## CFA with Lavaan package
library(lavaan)

Q.model <-'FACTOR1 =~ Q1 + Q2 + Q3 + Q4 + Q5 + Q6 + Q7 + Q8 + Q9
FACTOR2 =~ Q10 + Q11 + Q12 + Q13 + Q14
FACTOR3 =~ Q15 + Q16 + Q17 + Q18
FACTOR4 =~ Q19 + Q20 + Q21'

fit <- sem(Q.model, data = data.df, std.lv=TRUE, missing="fiml")
summary(fit, fit.measures=TRUE)
```

Results of Table 14 show that the presented model has a $X^2 = 593.222$, $X^2/df = 3.24$ with $p < 0.0001$. Additionally, CFI=0.879, TLI=0.861, RMSEA=0.105.This means that the model is acceptable, but it needs some adjustments.

In this sense, variables with less contribution to its factor are removed from the model: Q5, Q10, Q11, Q19.

The R code used to analyze the measurement model with the first adjustment is:

```
#Adjusted Model
library(lavaan)

Q.model.2 <-'FACTOR1 =~ Q1 + Q2 + Q3 + Q4 + Q6 + Q7 + Q8 + Q9
FACTOR2 =~ Q12 + Q13 + Q14
FACTOR3 =~ Q15 + Q16 + Q17 + Q18
FACTOR4 =~ Q20 + Q21'

fit <- sem(Q.model.2, data = data.df)
summary(fit, fit.measures=TRUE)
```

Table 15 in appendix present the results of the adjustment proposed. This model presents the following statistics: $X^2 = 386.521$, $X^2/df = 3.42$ with $p < 0.0001$. Additionally, CFI=0.900, TLI=0.879, RMSEA=0.109.

Table 3 summarize the comparison between these two models:

*Table 3. Comparison between models*

| Statistic | Values | | Fit of the Measurement Model | Fit of the Adjusted Measurement Model |
|---|---|---|---|---|
| $X^2/df$ | > 5<br>]2, 5]<br>]1, 2]<br>~1 | Bad<br>Suffering<br>Good<br>Very good | 3.24 | 3.42 |
| CFI<br>TLI | < 0.8<br>[0.8, 0.9[<br>[0.9, 0.95[<br>≥ 0.95 | Bad<br>Suffering<br>Good<br>Very good | 0.879<br>0.861 | 0.900<br>0.879 |
| RMSEA (I.C. 90%) e p-value (H0: RMSEA ≤ 0.05) | > 0.10<br>]0.08, 0.1]<br>]0.05, 0.08]<br>≤ 0.05 | Unacceptable<br>Mediocre<br>Good<br>Very good | 0.105 | 0.109 |

Measures obtained in both models allow us to conclude that the first model is better than the second, since it has better $X^2/df$ and RMSEA. Figure 4 represents the best measurement model analyzed.

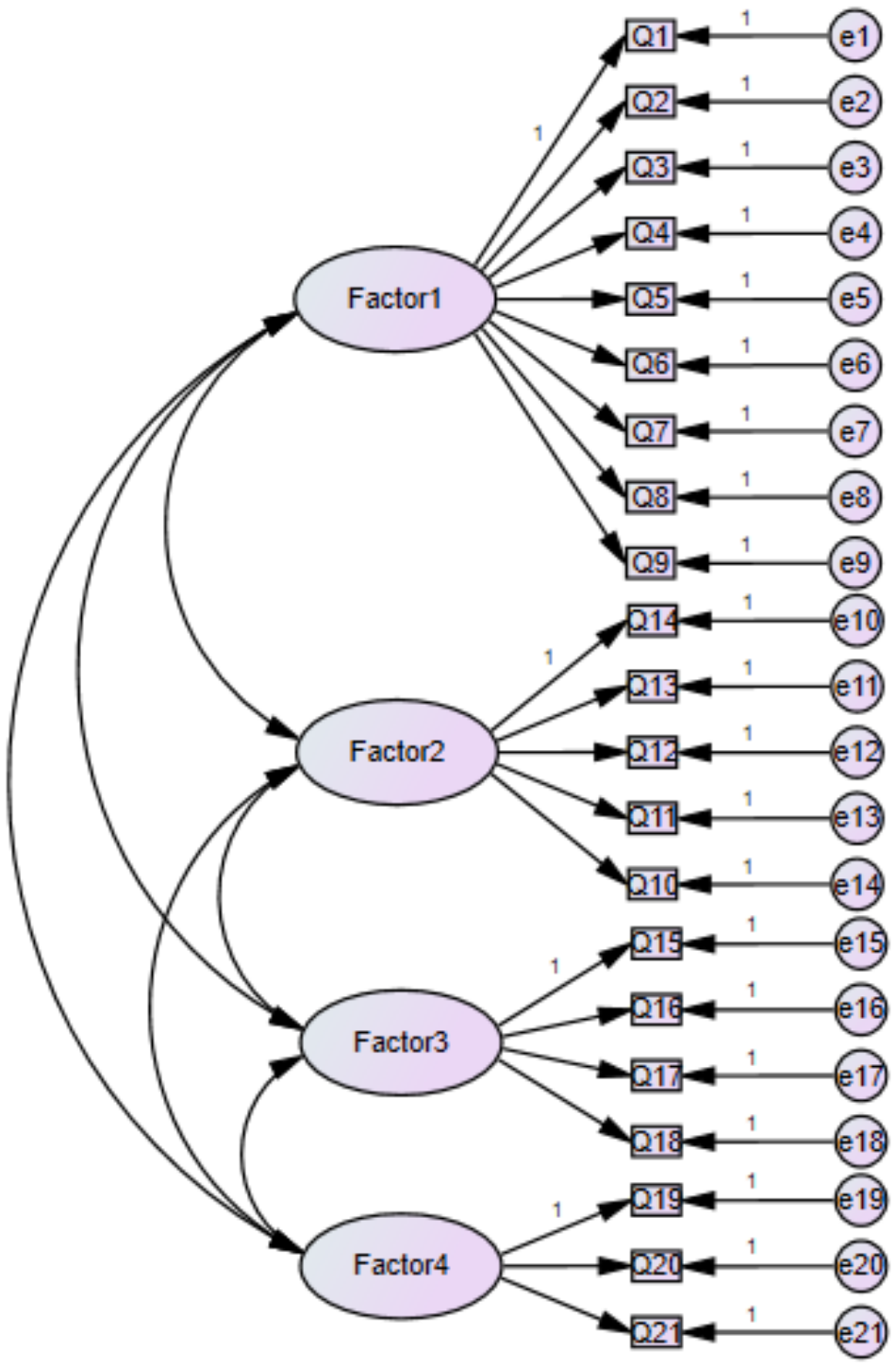

*Figure 4. Measurement model in study*

However, the obtained values indicates that both models are unsatisfactory and, therefore, some better adjustments should be performed. In this step, the researcher should explore several changes in the model (for example, associate correlated errors or remove variables) for to find the best model. The best model should have $X^2/df$ near to 1, CFI and TLI higher than 0.95 and RMSEA lower than 0.05.

**Structural model specification**

Structural model specification consists on the formal "drawing" of the measurement model, which reflects, a priori, the assumptions about the measurement model. Thus, the best measurement model with some directional influences should be explored in order to understand the influence of one factor in another factor. The model to be tested is represented in Figure 5.

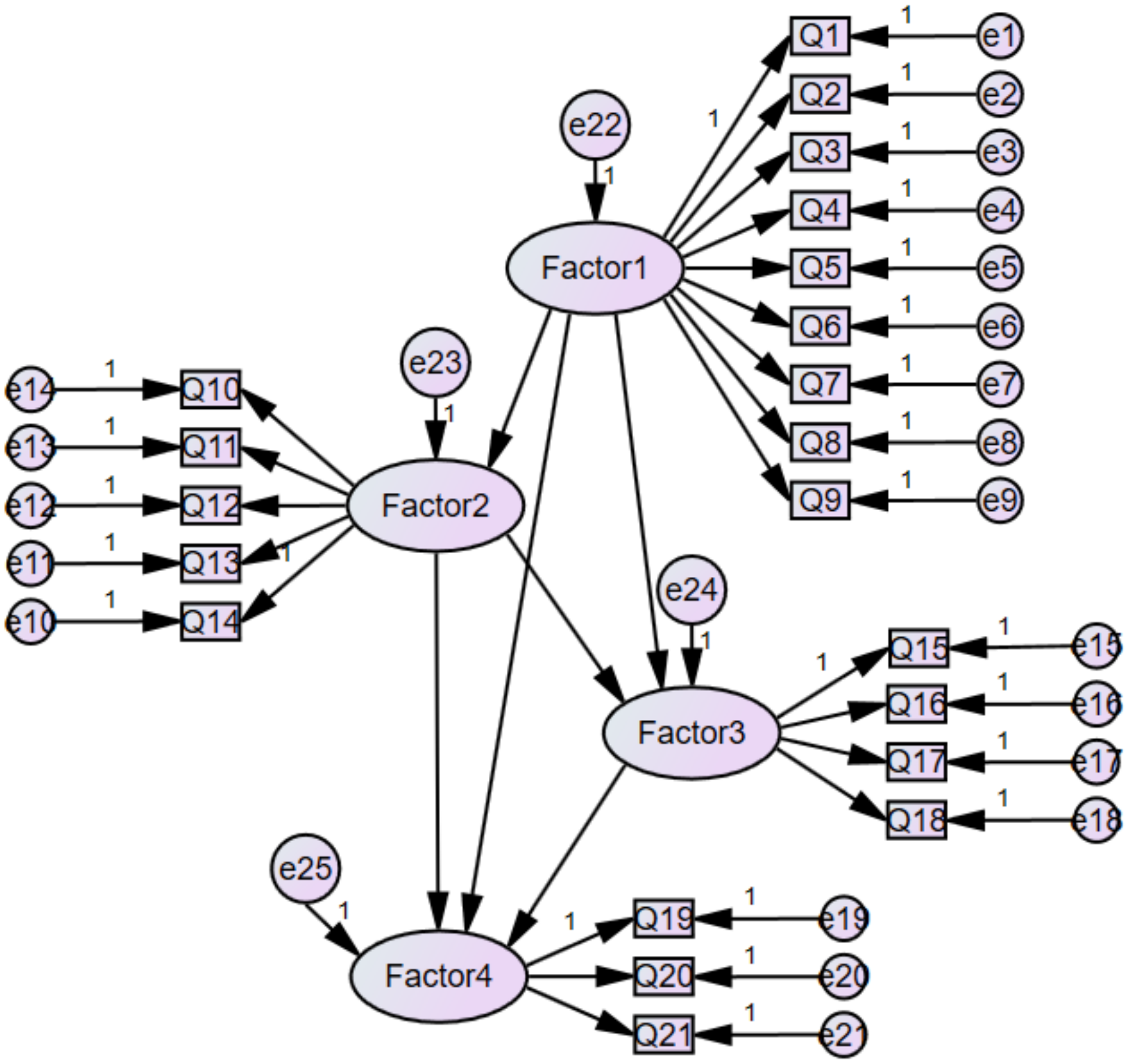

*Figure 5. Mental structural model*

R code used to perform the structural model is:

```
#with sem
library(sem)
cfa.model.2<-specifyModel("adjustedmodel.txt")
cfaOut.2<-sem(cfa.model.2,S=dataCov,N=204)
summary(cfaOut.2)
```

Table 16 presents the text file used as "adjusted model" introduced in the "specifyModel" function of the SEM R package. The output presented in Table 17 in appendix presents the result of the structural model specification. $X^2/df = 3.23$, which means it is very similar to the correspondent measurement model. In the end of the output are presented the following results:

```
        Estimate    Std Error   z value     Pr(>|z
F1F2DIR 0.407736694 0.083275769  4.89622252 9.7696435e-07 F2 <--- F1
F1F3DIR 0.902724046 0.112598958  8.01716163 1.0821668e-15 F3 <--- F1
F1F4DIR 0.520806176 0.131646222  3.95610423 7.6181932e-05 F4 <--- F1
F2F3DIR 0.034731466 0.082356936  0.42171878 6.7323029e-01 F3 <--- F2
F2F4DIR 0.176535492 0.088746145  1.98921871 4.6677067e-02 F4 <--- F2
F3F4DIR 0.286376634 0.093034567  3.07817453 2.0827290e-03 F4 <--- F3
```

Based on the given values, it is possible to conclude that F1→F2, F1→F3, F1→F4 and F3→F4 are statistically significant, which means that the first factor have a significant influence in the second factor. On the opposite, F2 does not statistically influence F3 ($p = 0.67 > 0.05$) because the null hypothesis of estimation equal to zero is rejected. By the same reason F2 does not influence F4 ($p = 0.0467 \approx 0.05$).
Thus, the structural model is represented in Figure 6.

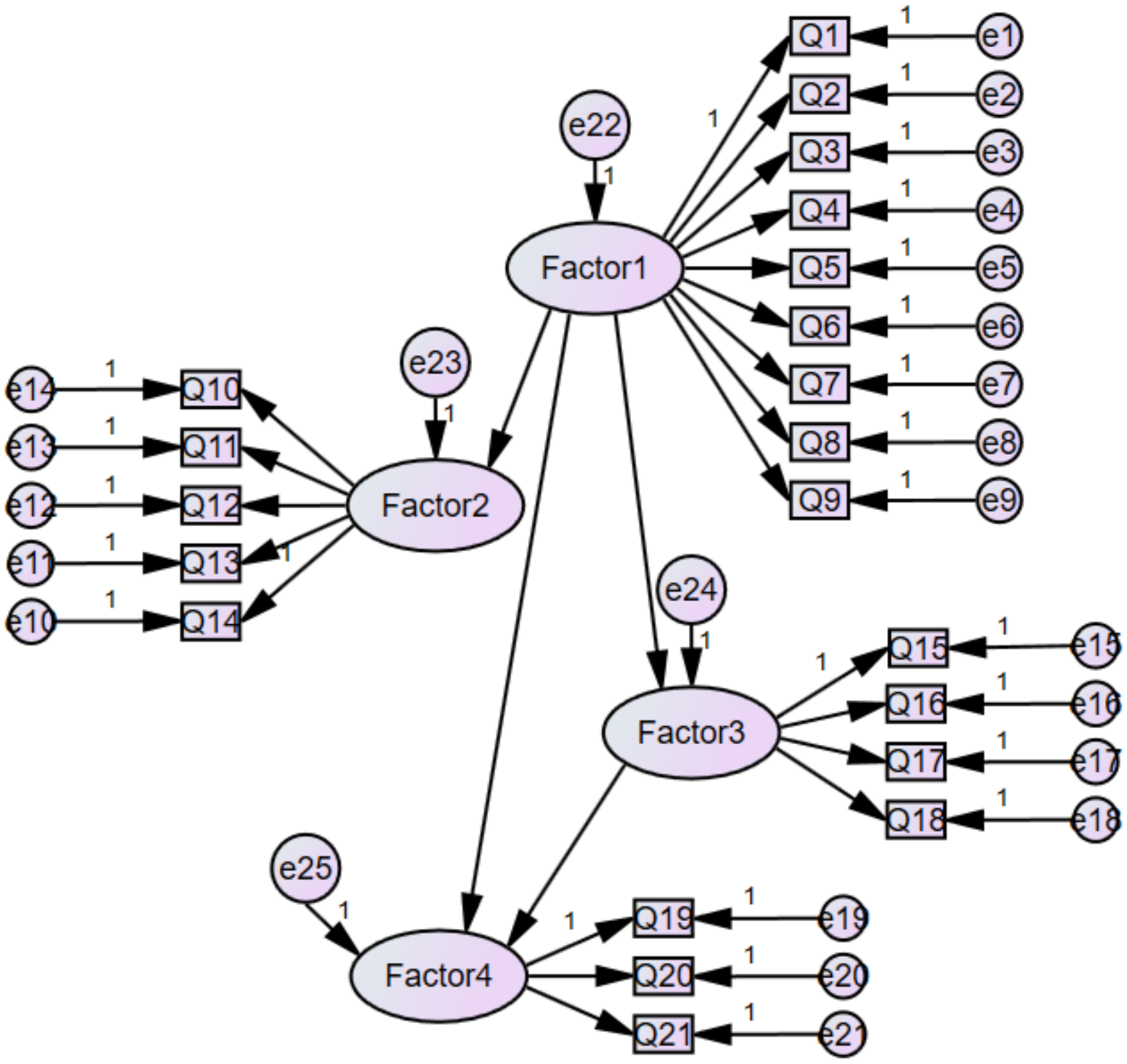

*Figure 6. Final structural model*

## SOLUTIONS AND RECOMMENDATIONS

As a recommendation in dealing with the eventual lack of knowledge of R language for statistical data analysis, specifically for the EFA, the reader might want to read the throughout R language application to statistics in (Sarmento & Costa, 2017).

## FUTURE RESEARCH DIRECTIONS

As future research direction, it can be useful for the researcher to explore other R packages available in CRAN repository, with the purpose to do EFA and CFA, or a more in depth research of both “sem” and “lavaan” R packages.

## CONCLUSION

Following the state of the art introduction, a throughout research template for CFA is written in this chapter. With the proposal of an example case study, the reader will hopefully appreciate the advantages within the use of CFA to explore his/her datasets regarding the hypothetical latent variables discovery and confirmation.

The authors presented a step by step approach to CFA, giving extreme importance to tasks like testing the fitness of the models and relevant statistical tests the reader might want to follow in their research. The reader is also introduced to the tasks he/she will have to do to explore even further the concept of CFA with their data.

## ACKNOWLEDGMENT

Rui Portocarrero Sarmento also gratefully acknowledges funding from FCT (Portuguese Foundation for Science and Technology) through a PhD grant (SFRH/BD/119108/2016). Vera Costa gratefully acknowledges funding from FCT (Portuguese Foundation for Science and Technology) through a Ph.D. grant (Ref. PD/BD/128065/2016). The authors want to thank also to the reviewers for the constructive reviews provided in the development of this publication.

## REFERENCES

Furr, M. (2011). Confirmatory Factor Analysis. In *Scale Construction and Psychometrics for Social and Personality Psychology* (pp. 1–30). http://doi.org/10.4135/9781412961288.n69

Gunzler, D. D., & Morris, N. (2016). *A Tutorial on Structural Equation Modeling for Analysis of Overlapping Symptoms in Co-occurring Conditions Using MPlus* (Vol. 34). http://doi.org/10.1002/sim.6541.A

Gutierrez, G. C. (2005). Modelo de Equações Estruturais. In *Estimação das escalas dos construtos capital social, capital cultural e capital econômico e análise do efeito escola nos dados do Peru-PISA 2000* (pp. 70–88). Retrieved from https://www.maxwell.vrac.puc-rio.br/7630/7630_7.PDF

Joseph F. Hair, J., Black, W. C., Babin, B. J., Anderson, R. E., & Tatham, R. L. (2009). *Análise Multivariada de Dados*.

Leach, C. W., van Zomeren, M., Zebel, S., Vliek, M. L. W., Pennekamp, S. F., Doosje, B., … Spears, R. (2008). Group-Level Self-Definition and Self-Investment: A Hierarchical (Multicomponent) Model of In-Group Identification. *Journal of Personality and Social Psychology*, *95*(1), 144–165. http://doi.org/10.1037/0022-3514.95.1.144

Lewis, T. F. (2017). Evidence regarding the internal structure: Confirmatory factor analysis.

*Measurement and Evaluation in Counseling and Development*, *50*(4), 239–247. http://doi.org/10.1080/07481756.2017.1336929

Maccallum, R. C., Browne, M. W., & Sugawara, H. M. (1996). Power analysis and determination of sample size for covariance structure modeling of fit involving a particular measure of model. *Psychological Methods*, *13*(2), 130–149. http://doi.org/10.1037/1082-989X.1.2.130

Malhotra, N., Hall, J., Shaw, M., & Oppenheim, P. (2007). *Essentials of Marketing Research, An Applied Orientation*.

Portela, D. M. P. (2012). *Contributo das Técnicas de Análise Fatorial para o Estudo do Programa "Ocupação Científica de Jovens nas Férias." Universidade aberta 2010*.

Sánchez, F. J. S. (1999). Capítulo III Metodología de la Investigación. In *Validez y fiabilidad de escalas*.

Sarmento, R., & Costa, V. (2017). *Comparative approaches to using R and Python for statistical data analysis*. *Comparative Approaches to Using R and Python for Statistical Data Analysis*. http://doi.org/10.4018/978-1-68318-016-6

Sarmento, R., & Costa, V. (2017). Factor Analysis. In *Comparative Approaches to Using R and Python for Statistical Data Analysis* (pp. 148–178). http://doi.org/10.4018/978-1-68318-016-6

Silva, R. P. A., Macêdo, L. C. B., & Silva, I. L. R. (2013). Avaliação das Características Psicométricas dos Questionários Utilizados nos Periódicos da Area Contábil: Um Estudo Longitudinal Compreendido no Período 2003-2012 . In *XX Congresso Brasileiro de Custos*.

Suhr, D. (2006). Exploratory or Confirmatory Factor Analysis? *Statistics and Data Analysis*, 1–17. http://doi.org/10.1002/da.20406

# APPENDIX

*Table 4*

| | |
|---|---|
| *Computer Code* | `head(data.df)` |

*Output*

```
  Q1 Q2 Q3 Q4 Q5 Q6 Q7 Q8 Q9 Q10 Q11 Q12 Q13 Q14 Q15 Q16 Q17 Q18 Q19
1  4  4  6  4  5  5  5  2  4   5   4   6   4   4   5   3   1   6   7
2  5  5  5  5  5  5  5  5  5   1   1   1   1   1   6   6   6   6   7
3  2  2  2  2  4  2  2  2  2   5   4   4   4   4   3   3   3   3   5
4  5  5  5  5  5  5  5  5  5   5   5   5   4   4   4   4   3   3   6
5  2  5  5  4  4  4  3  4  2   5   4   4   3   3   5   4   5   3   5
6  3  5  5  5  4  4  3  3  6   5   4   4   4   4   4   3   3   4   5
  Q20 Q21
1   2   5
2   5   5
3   3   2
4   6   5
5   4   2
6   2   3
```

*Table 5*

*Computer Code*

```
### Descriptive analysis of Q1 to Q21 variables
# Identification of the variables used in factor analysis
survey<-data.df[, paste("Q", 1:21, sep="")]

# Descriptive analysis for each variable
summary(survey)
```

<table>
<tr><td>Output</td><td><pre>
       Q1              Q2              Q3              Q4       
 Min.   :1.000   Min.   :1.000   Min.   :1.000   Min.   :1.000  
 1st Qu.:3.000   1st Qu.:3.000   1st Qu.:4.000   1st Qu.:4.000  
 Median :5.000   Median :5.000   Median :5.000   Median :5.000  
 Mean   :4.392   Mean   :4.471   Mean   :4.892   Mean   :4.608  
 3rd Qu.:6.000   3rd Qu.:6.000   3rd Qu.:6.000   3rd Qu.:6.000  
 Max.   :7.000   Max.   :7.000   Max.   :7.000   Max.   :7.000  
       Q5              Q6             Q7              Q8       
 Min.   :1.000   Min.   :1.00   Min.   :1.000   Min.   :1.000  
 1st Qu.:4.000   1st Qu.:4.00   1st Qu.:3.000   1st Qu.:2.000  
 Median :5.000   Median :5.00   Median :5.000   Median :4.000  
 Mean   :5.054   Mean   :5.01   Mean   :4.505   Mean   :3.603  
 3rd Qu.:6.000   3rd Qu.:6.00   3rd Qu.:6.000   3rd Qu.:5.000  
 Max.   :7.000   Max.   :7.00   Max.   :7.000   Max.   :7.000  
       Q9             Q10             Q11             Q12       
 Min.   :1.000   Min.   :1.000   Min.   :1.000   Min.   :1.000  
 1st Qu.:3.000   1st Qu.:5.000   1st Qu.:4.000   1st Qu.:4.000  
 Median :4.000   Median :5.000   Median :4.000   Median :4.000  
 Mean   :4.176   Mean   :4.789   Mean   :4.015   Mean   :3.936  
 3rd Qu.:5.000   3rd Qu.:5.000   3rd Qu.:4.000   3rd Qu.:4.000  
 Max.   :7.000   Max.   :7.000   Max.   :7.000   Max.   :7.000  
      Q13             Q14             Q15             Q16       
 Min.   :1.000   Min.   :1.000   Min.   :1.000   Min.   :1.000  
 1st Qu.:4.000   1st Qu.:4.000   1st Qu.:3.000   1st Qu.:3.000  
 Median :4.000   Median :4.000   Median :5.000   Median :4.000  
 Mean   :3.975   Mean   :3.951   Mean   :4.578   Mean   :4.328  
 3rd Qu.:4.000   3rd Qu.:4.000   3rd Qu.:6.000   3rd Qu.:6.000  
 Max.   :7.000   Max.   :7.000   Max.   :7.000   Max.   :7.000  
      Q17             Q18             Q19             Q20       
 Min.   :1.000   Min.   :1.000   Min.   :1.000   Min.   :1.000  
 1st Qu.:3.000   1st Qu.:3.000   1st Qu.:5.000   1st Qu.:4.000  
 Median :4.000   Median :5.000   Median :6.000   Median :5.000  
 Mean   :3.941   Mean   :4.461   Mean   :5.627   Mean   :4.721  
 3rd Qu.:5.000   3rd Qu.:6.000   3rd Qu.:6.000   3rd Qu.:6.000  
 Max.   :7.000   Max.   :7.000   Max.   :7.000   Max.   :7.000  
      Q21       
 Min.   :1.000  
 1st Qu.:3.000  
 Median :5.000  
 Mean   :4.529  
 3rd Qu.:6.000  
 Max.   :7.000  
</pre></td></tr>
</table>

*Table 6*

<table>
<tr><td>Computer Code</td><td><pre>
### Correlation between variables Q1 to Q10
correlation <- cor(survey, method="spearman")
correlation
</pre></td></tr>
</table>

*Output*

| | Q1 | Q2 | Q3 | Q4 | Q5 | Q6 |
|---|---|---|---|---|---|---|
| Q1 | 1.0000000 | 0.8678069 | 0.7103314 | 0.6795034 | 0.6317822 | 0.7219436 |
| Q2 | 0.8678069 | 1.0000000 | 0.7140347 | 0.6607190 | 0.6411241 | 0.6852318 |
| Q3 | 0.7103314 | 0.7140347 | 1.0000000 | 0.6923580 | 0.6288720 | 0.7034031 |
| Q4 | 0.6795034 | 0.6607190 | 0.6923580 | 1.0000000 | 0.5815030 | 0.6920358 |
| Q5 | 0.6317822 | 0.6411241 | 0.6288720 | 0.5815030 | 1.0000000 | 0.8010300 |
| Q6 | 0.7219436 | 0.6852318 | 0.7034031 | 0.6920358 | 0.8010300 | 1.0000000 |
| Q7 | 0.7019616 | 0.7270406 | 0.6654539 | 0.6484587 | 0.5664464 | 0.5753212 |
| Q8 | 0.7181340 | 0.6549639 | 0.5127171 | 0.5824695 | 0.4838123 | 0.5698544 |
| Q9 | 0.6789200 | 0.6922414 | 0.6301799 | 0.6198808 | 0.4999616 | 0.5347266 |
| Q10 | 0.2734871 | 0.2466369 | 0.3735019 | 0.2986515 | 0.3344236 | 0.3587571 |
| Q11 | 0.1983869 | 0.2414378 | 0.2923634 | 0.2050208 | 0.2923783 | 0.2222434 |
| Q12 | 0.2958467 | 0.2768418 | 0.2966513 | 0.2690579 | 0.2634822 | 0.2925383 |
| Q13 | 0.2459464 | 0.2949535 | 0.2422306 | 0.2187331 | 0.3068287 | 0.2190018 |
| Q14 | 0.2589249 | 0.2207151 | 0.3095463 | 0.2407727 | 0.3282584 | 0.2953405 |
| Q15 | 0.5379912 | 0.5317333 | 0.5411050 | 0.4914183 | 0.4707387 | 0.5171374 |
| Q16 | 0.5548849 | 0.4641767 | 0.5002715 | 0.4551067 | 0.4373705 | 0.5076297 |
| Q17 | 0.4921372 | 0.4872918 | 0.3837584 | 0.4062902 | 0.3649162 | 0.3993855 |
| Q18 | 0.5252167 | 0.4358864 | 0.5149534 | 0.4575959 | 0.4047356 | 0.4785591 |
| Q19 | 0.3469913 | 0.4003287 | 0.4503138 | 0.4144919 | 0.4303146 | 0.3974906 |
| Q20 | 0.3451128 | 0.3693582 | 0.4061743 | 0.4837740 | 0.3534774 | 0.3804239 |
| Q21 | 0.4577953 | 0.5093471 | 0.5774302 | 0.6110928 | 0.4573397 | 0.4600055 |

| | Q7 | Q8 | Q9 | Q10 | Q11 | Q12 |
|---|---|---|---|---|---|---|
| Q1 | 0.7019616 | 0.7181340 | 0.6789200 | 0.2734871 | 0.1983869 | 0.29584665 |
| Q2 | 0.7270406 | 0.6549639 | 0.6922414 | 0.2466369 | 0.2414378 | 0.27684175 |
| Q3 | 0.6654539 | 0.5127171 | 0.6301799 | 0.3735019 | 0.2923634 | 0.29665134 |
| Q4 | 0.6484587 | 0.5824695 | 0.6198808 | 0.2986515 | 0.2050208 | 0.26905787 |
| Q5 | 0.5664464 | 0.4838123 | 0.4999616 | 0.3344236 | 0.2923783 | 0.26348222 |
| Q6 | 0.5753212 | 0.5698544 | 0.5347266 | 0.3587571 | 0.2222434 | 0.29253828 |
| Q7 | 1.0000000 | 0.6635452 | 0.7492209 | 0.2954756 | 0.2747056 | 0.27690266 |
| Q8 | 0.6635452 | 1.0000000 | 0.5986723 | 0.1978725 | 0.1283247 | 0.18568166 |
| Q9 | 0.7492209 | 0.5986723 | 1.0000000 | 0.2591136 | 0.3143535 | 0.27667083 |
| Q10 | 0.2954756 | 0.1978725 | 0.2591136 | 1.0000000 | 0.5729275 | 0.50225663 |
| Q11 | 0.2747056 | 0.1283247 | 0.3143535 | 0.5729275 | 1.0000000 | 0.60452377 |
| Q12 | 0.2769027 | 0.1856817 | 0.2766708 | 0.5022566 | 0.6045238 | 1.00000000 |
| Q13 | 0.2882333 | 0.1373405 | 0.2526130 | 0.5044710 | 0.6843456 | 0.58692135 |
| Q14 | 0.2120628 | 0.1930969 | 0.1777866 | 0.6030657 | 0.3998505 | 0.50390205 |
| Q15 | 0.6189304 | 0.5194451 | 0.5336565 | 0.2718297 | 0.2406985 | 0.22294611 |
| Q16 | 0.5198095 | 0.5329469 | 0.4366183 | 0.2381233 | 0.2005993 | 0.22311055 |
| Q17 | 0.5981843 | 0.5589763 | 0.5110475 | 0.1661340 | 0.1658558 | 0.09045817 |
| Q18 | 0.5364725 | 0.4363748 | 0.4710596 | 0.2670542 | 0.1828065 | 0.23007770 |
| Q19 | 0.5076996 | 0.2796218 | 0.4718240 | 0.2269147 | 0.2086547 | 0.22728771 |
| Q20 | 0.4736844 | 0.3729992 | 0.4012173 | 0.2475175 | 0.1376753 | 0.25300100 |
| Q21 | 0.5294104 | 0.3972187 | 0.4783270 | 0.2596550 | 0.2367676 | 0.20166012 |

| | Q13 | Q14 | Q15 | Q16 | Q17 | Q18 |
|---|---|---|---|---|---|---|
| Q1 | 0.2459464 | 0.25892487 | 0.5379912 | 0.5548849 | 0.49213718 | 0.5252167 |
| Q2 | 0.2949535 | 0.22071512 | 0.5317333 | 0.4641767 | 0.48729176 | 0.4358864 |
| Q3 | 0.2422306 | 0.30954633 | 0.5411050 | 0.5002715 | 0.38375839 | 0.5149534 |
| Q4 | 0.2187331 | 0.24077267 | 0.4914183 | 0.4551067 | 0.40629024 | 0.4575959 |
| Q5 | 0.3068287 | 0.32825843 | 0.4707387 | 0.4373705 | 0.36491621 | 0.4047356 |
| Q6 | 0.2190018 | 0.29534045 | 0.5171374 | 0.5076297 | 0.39938546 | 0.4785591 |
| Q7 | 0.2882333 | 0.21206282 | 0.6189304 | 0.5198095 | 0.59818431 | 0.5364725 |
| Q8 | 0.1373405 | 0.19309690 | 0.5194451 | 0.5329469 | 0.55897635 | 0.4363748 |
| Q9 | 0.2526130 | 0.17778659 | 0.5336565 | 0.4366183 | 0.51104746 | 0.4710596 |
| Q10 | 0.5044710 | 0.60306566 | 0.2718297 | 0.2381233 | 0.16613402 | 0.2670542 |
| Q11 | 0.6843456 | 0.39985053 | 0.2406985 | 0.2005993 | 0.16585577 | 0.1828065 |
| Q12 | 0.5869213 | 0.50390205 | 0.2229461 | 0.2231106 | 0.09045817 | 0.2300777 |
| Q13 | 1.0000000 | 0.60438408 | 0.2859497 | 0.2460159 | 0.16990677 | 0.2287251 |
| Q14 | 0.6043841 | 1.00000000 | 0.2157779 | 0.2292658 | 0.09505105 | 0.2101240 |
| Q15 | 0.2859497 | 0.21577791 | 1.0000000 | 0.7628373 | 0.70033301 | 0.6967244 |
| Q16 | 0.2460159 | 0.22926581 | 0.7628373 | 1.0000000 | 0.71835003 | 0.8041077 |
| Q17 | 0.1699068 | 0.09505105 | 0.7003330 | 0.7183500 | 1.00000000 | 0.6313257 |
| Q18 | 0.2287251 | 0.21012404 | 0.6967244 | 0.8041077 | 0.63132571 | 1.0000000 |

| | |
|---|---|
| | ```<br>Q19 0.1966188 0.16358945 0.4908673 0.3844242 0.38799527 0.4437708<br>Q20 0.2015674 0.30205409 0.4592085 0.3880153 0.35790683 0.4098546<br>Q21 0.2582761 0.26929830 0.4930299 0.3864878 0.30046152 0.3983685<br>          Q19       Q20       Q21<br>Q1  0.3469913 0.3451128 0.4577953<br>Q2  0.4003287 0.3693582 0.5093471<br>Q3  0.4503138 0.4061743 0.5774302<br>Q4  0.4144919 0.4837740 0.6110928<br>Q5  0.4303146 0.3534774 0.4573397<br>Q6  0.3974906 0.3804239 0.4600055<br>Q7  0.5076996 0.4736844 0.5294104<br>Q8  0.2796218 0.3729992 0.3972187<br>Q9  0.4718240 0.4012173 0.4783270<br>Q10 0.2269147 0.2475175 0.2596550<br>Q11 0.2086547 0.1376753 0.2367676<br>Q12 0.2272877 0.2530010 0.2016601<br>Q13 0.1966188 0.2015674 0.2582761<br>Q14 0.1635894 0.3020541 0.2692983<br>Q15 0.4908673 0.4592085 0.4930299<br>Q16 0.3844242 0.3880153 0.3864878<br>Q17 0.3879953 0.3579068 0.3004615<br>Q18 0.4437708 0.4098546 0.3983685<br>Q19 1.0000000 0.6595408 0.5836302<br>Q20 0.6595408 1.0000000 0.6988842<br>Q21 0.5836302 0.6988842 1.0000000<br>``` |

*Table 7*

| | |
|---|---|
| *Computer Code* | ### Kaiser criterion<br>library (psych)<br>eigen(correlation) |

*Output*

```
$values
 [1] 9.7900202 2.5793759 1.4564320 1.3194869 0.8433026 0.6650069
 [7] 0.5292425 0.5214856 0.4940456 0.4299674 0.3260301 0.2866790
[13] 0.2835303 0.2498692 0.2411731 0.2402759 0.1989649 0.1578476
[19] 0.1475062 0.1396329 0.1001252

$vectors
            [,1]         [,2]          [,3]         [,4]         [,5]
 [1,] -0.2644054 -0.103713896  0.2518552377  0.172795263 -0.005945114
 [2,] -0.2611277 -0.090076282  0.2770726345  0.086538519 -0.147572521
 [3,] -0.2585339 -0.028900816  0.2008702172 -0.043383347  0.106441837
 [4,] -0.2499055 -0.075651954  0.2141669207 -0.115359617  0.052091520
 [5,] -0.2364852  0.008149364  0.2429626716 -0.001248406  0.293954529
 [6,] -0.2516333 -0.045303963  0.2580419408  0.042526965  0.354572006
 [7,] -0.2660270 -0.092468041  0.0521092567  0.019592551 -0.296973922
 [8,] -0.2295898 -0.159379164  0.1184747655  0.181643877 -0.095004775
 [9,] -0.2463834 -0.076653178  0.1169753726  0.021776400 -0.425368280
[10,] -0.1512541  0.390735969 -0.0138765940  0.034526475  0.205514113
[11,] -0.1330244  0.432986536 -0.0478578366  0.104100028 -0.348644414
[12,] -0.1406055  0.409655680 -0.0006854575  0.049422968 -0.150662571
[13,] -0.1406522  0.444233530 -0.0889495608  0.095818323 -0.157344511
[14,] -0.1348379  0.401696915 -0.0127032005 -0.020144984  0.379846551
[15,] -0.2441944 -0.098573894 -0.3320234666  0.106262161  0.036518088
[16,] -0.2301132 -0.109858480 -0.3786076579  0.245727258  0.216310533
[17,] -0.2098132 -0.180087909 -0.3516474497  0.255591359 -0.127028060
[18,] -0.2245834 -0.098232755 -0.3757778500  0.155086662  0.191295621
[19,] -0.1959508 -0.037841987 -0.2119704530 -0.459447396 -0.128857335
[20,] -0.1949735 -0.016517763 -0.2072220177 -0.534403190  0.028653871
[21,] -0.2177550 -0.019471863 -0.0255831386 -0.479594335 -0.006881026
             [,6]        [,7]         [,8]        [,9]       [,10]
 [1,]  0.12457723  0.12086601 -0.118768110 -0.01823734  0.15932051
 [2,]  0.03427313  0.16971433  0.047490641  0.07576608  0.19926305
 [3,] -0.17977751 -0.22595406 -0.233619259  0.22788288  0.26743042
 [4,]  0.06744192 -0.11963141 -0.256579984  0.21017011 -0.28920798
 [5,] -0.37310128  0.24017152  0.382700039 -0.14317959 -0.11289798
 [6,] -0.24101204  0.08285398  0.024299674 -0.17948715 -0.20266524
 [7,]  0.08103093 -0.12872862  0.124290510 -0.05541144  0.15837814
 [8,]  0.47775754  0.01668529  0.107977141 -0.19821434 -0.37205916
 [9,]  0.01267051 -0.21162835  0.002971529 -0.05590552  0.27317883
[10,]  0.03702361 -0.70453367  0.101163017 -0.16089039 -0.07411916
[11,] -0.32486749 -0.10104438  0.060553372  0.14693651 -0.34666309
[12,]  0.02590782  0.28173577 -0.575316460 -0.42176319 -0.09688799
[13,]  0.01979179  0.38392878  0.228562964  0.33206926  0.10183554
[14,]  0.45620299  0.05651720  0.193552439  0.01633625  0.37280756
[15,] -0.08652328  0.02187092  0.065140250  0.14506606 -0.02830146
[16,] -0.02655728  0.10256753 -0.156758306  0.08532211 -0.07726191
[17,]  0.06958386 -0.06310267  0.307974476 -0.09869065 -0.13805103
[18,] -0.12311628 -0.04439791 -0.330393799  0.05092845  0.21301898
[19,] -0.29570884  0.03802657  0.141413670 -0.39845674  0.25619467
[20,]  0.27749397  0.09941307 -0.008902055 -0.16402062 -0.16761456
[21,]  0.07372829  0.01447918 -0.039580405  0.48723586 -0.19607492
             [,11]       [,12]        [,13]        [,14]       [,15]
 [1,] -0.301998558 -0.04463061 -0.283638258  0.042376709 -0.15288710
 [2,] -0.360204858  0.21131223 -0.291850346  0.101984107 -0.07997436
 [3,] -0.162270494  0.07853730  0.297078197  0.304349422  0.34653975
 [4,]  0.580259685  0.20253097 -0.288924577  0.030301239 -0.10438674
 [5,]  0.127449838 -0.14100786  0.147175232 -0.094366695  0.14014376
 [6,]  0.093511178  0.03184000  0.008482818  0.011602852 -0.13893157
 [7,]  0.164256329  0.14336166  0.091667271 -0.418887810  0.59954284
 [8,] -0.192320076 -0.34645611  0.166938375 -0.172633736  0.01351558
 [9,]  0.345430864 -0.28767603  0.246996927  0.089568995 -0.45762803
[10,] -0.202264477  0.11449160 -0.253573396 -0.215948210 -0.10477723
[11,] -0.129609043 -0.27599167  0.082135200  0.262433165  0.06061117
```

```
[12,]  0.003695023  0.24173037  0.168728106 -0.004684864  0.06215031
[13,]  0.142375851  0.01506302 -0.285592381 -0.236650246 -0.01921524
[14,]  0.165345532 -0.05871857  0.231123699  0.245402983 -0.02428839
[15,] -0.112484228  0.46394183  0.429988842 -0.315225846 -0.37717986
[16,] -0.075197088 -0.16081363  0.022030854  0.104339422 -0.01514057
[17,]  0.129668968  0.29577521 -0.159298792  0.517501290  0.15893955
[18,]  0.104352860 -0.41641187 -0.236833164 -0.212300904  0.12542539
[19,] -0.087662801 -0.04664906 -0.165422502  0.011804632 -0.13172318
[20,] -0.016641223 -0.02312072 -0.071130845  0.120218987  0.10182202
[21,] -0.220075474 -0.07645795  0.099330401 -0.049467588 -0.02374009
            [,16]        [,17]        [,18]        [,19]       [,20]
 [1,]  0.06958498  0.030708684  0.242956159  0.035516681 -0.08719873
 [2,]  0.14282045  0.019071124  0.138317064 -0.055775219  0.09914793
 [3,] -0.22574777 -0.313897951 -0.278360761  0.080662819  0.20907252
 [4,] -0.30677818  0.006454483  0.257821790  0.061848405  0.13502317
 [5,]  0.20005690  0.214901438  0.112670807  0.258389479  0.40913761
 [6,]  0.08131217 -0.197424677 -0.295840529 -0.368455773 -0.54769146
 [7,]  0.06327764  0.054165621  0.182180151  0.005760193 -0.36273949
 [8,] -0.38450035 -0.076367055 -0.166379979 -0.081882085  0.19407008
 [9,]  0.28101870  0.010167834 -0.177904596  0.150773645 -0.01856988
[10,]  0.10932535  0.053933731 -0.120668837  0.178105364  0.06281266
[11,] -0.03020887 -0.121359160  0.402205604 -0.201561449 -0.09297739
[12,]  0.07229733  0.256458739 -0.135444752  0.035125269  0.08665805
[13,] -0.11212039 -0.256719377 -0.402221608  0.109164493  0.03479825
[14,] -0.11363812  0.138175325  0.254950027 -0.177546426 -0.09771485
[15,] -0.02236043 -0.163476248  0.202455153 -0.153112955  0.14724912
[16,] -0.08539113  0.017059638  0.077378864  0.638298070 -0.36890373
[17,]  0.07810167  0.224852868 -0.242713916 -0.144244085  0.09150319
[18,]  0.14515625  0.043960423  0.006328784 -0.413362078  0.25003041
[19,] -0.53735037  0.083868833  0.027280569 -0.009488860 -0.05359209
[20,]  0.41789033 -0.510520905  0.097157407  0.086062147  0.08282710
[21,]  0.10158302  0.538993200 -0.218125705 -0.078989914 -0.12010117
             [,21]
 [1,]  0.695388552
 [2,] -0.640917867
 [3,]  0.081799690
 [4,] -0.041637997
 [5,]  0.050819930
 [6,] -0.059831083
 [7,]  0.007236728
 [8,] -0.085606648
 [9,] -0.035775801
[10,] -0.021348293
[11,] -0.004806558
[12,] -0.008693214
[13,]  0.092354400
[14,] -0.066957178
[15,]  0.077323707
[16,] -0.187653522
[17,]  0.135894453
[18,] -0.078678282
[19,]  0.007140874
[20,]  0.027189683
[21,]  0.046938987
```

*Table 8*

**Computer Code**

```
### Principal Component method
library (psych)
principal(correlation,nfactors=4, rotate="none")
```

**Output**

```
Principal Components Analysis
Call: principal(r = correlation, nfactors = 4, rotate = "none")
Standardized loadings (pattern matrix) based upon correlation matrix
     PC1   PC2   PC3   PC4   h2   u2 com
Q1  0.83 -0.17 -0.30 -0.20 0.84 0.16 1.5
Q2  0.82 -0.14 -0.33 -0.10 0.81 0.19 1.4
Q3  0.81 -0.05 -0.24  0.05 0.72 0.28 1.2
Q4  0.78 -0.12 -0.26  0.13 0.71 0.29 1.3
Q5  0.74  0.01 -0.29  0.00 0.63 0.37 1.3
Q6  0.79 -0.07 -0.31 -0.05 0.72 0.28 1.3
Q7  0.83 -0.15 -0.06 -0.02 0.72 0.28 1.1
Q8  0.72 -0.26 -0.14 -0.21 0.65 0.35 1.5
Q9  0.77 -0.12 -0.14 -0.03 0.63 0.37 1.1
Q10 0.47  0.63  0.02 -0.04 0.62 0.38 1.9
Q11 0.42  0.70  0.06 -0.12 0.67 0.33 1.7
Q12 0.44  0.66  0.00 -0.06 0.63 0.37 1.8
Q13 0.44  0.71  0.11 -0.11 0.73 0.27 1.8
Q14 0.42  0.65  0.02  0.02 0.59 0.41 1.7
Q15 0.76 -0.16  0.40 -0.12 0.78 0.22 1.7
Q16 0.72 -0.18  0.46 -0.28 0.84 0.16 2.2
Q17 0.66 -0.29  0.42 -0.29 0.78 0.22 2.6
Q18 0.70 -0.16  0.45 -0.18 0.76 0.24 2.0
Q19 0.61 -0.06  0.26  0.53 0.72 0.28 2.3
Q20 0.61 -0.03  0.25  0.61 0.81 0.19 2.3
Q21 0.68 -0.03  0.03  0.55 0.77 0.23 1.9

                       PC1  PC2  PC3  PC4
SS loadings           9.79 2.58 1.46 1.32
Proportion Var        0.47 0.12 0.07 0.06
Cumulative Var        0.47 0.59 0.66 0.72
Proportion Explained  0.65 0.17 0.10 0.09
Cumulative Proportion 0.65 0.82 0.91 1.00

Mean item complexity =  1.7
Test of the hypothesis that 4 components are sufficient.

The root mean square of the residuals (RMSR) is  0.05

Fit based upon off diagonal values = 0.99

Importance of components:
                          PC1     PC2     PC3     PC4     PC5     PC6
Standard deviation     5.2041 2.19297 1.93647 1.69202 1.44830 1.21755
Proportion of Variance 0.5257 0.09336 0.07279 0.05558 0.04072 0.02878
Cumulative Proportion  0.5257 0.61909 0.69189 0.74746 0.78818 0.81696
                           PC7     PC8     PC9    PC10    PC11
Standard deviation     1.14315 1.03692 1.03248 0.91380 0.87686
Proportion of Variance 0.02537 0.02087 0.02069 0.01621 0.01493
Cumulative Proportion  0.84232 0.86320 0.88389 0.90010 0.91503
                          PC12    PC13    PC14    PC15    PC16
Standard deviation     0.87483 0.82893 0.75825 0.68115 0.66079
Proportion of Variance 0.01486 0.01334 0.01116 0.00901 0.00848
Cumulative Proportion  0.92988 0.94322 0.95438 0.96339 0.97187
                          PC17    PC18    PC19   PC20    PC21
Standard deviation     0.62693 0.59443 0.54742 0.5125 0.37496
Proportion of Variance 0.00763 0.00686 0.00582 0.0051 0.00273
Cumulative Proportion  0.97950 0.98635 0.99217 0.9973 1.00000
```

*Table 9*

| *Computer Code* | *Output* |
|---|---|

**Computer Code**

```
### Principal Component method with varimax rotation
library (psych)
principal(correlation,nfactors=4, rotate="varimax")
```

**Output**

```
Principal Components Analysis
Call: principal(r = correlation, nfactors = 4, rotate = "varimax")
Standardized loadings (pattern matrix) based upon correlation matrix
     PC1  PC2  PC3  PC4   h2   u2 com
Q1  0.86 0.13 0.30 0.06 0.84 0.16 1.3
Q2  0.85 0.14 0.22 0.13 0.81 0.19 1.2
Q3  0.74 0.22 0.19 0.28 0.72 0.28 1.6
Q4  0.74 0.13 0.16 0.35 0.71 0.29 1.6
Q5  0.72 0.25 0.13 0.20 0.63 0.37 1.5
Q6  0.79 0.19 0.18 0.17 0.72 0.28 1.3
Q7  0.68 0.15 0.39 0.28 0.72 0.28 2.1
Q8  0.70 0.02 0.39 0.06 0.65 0.35 1.6
Q9  0.68 0.15 0.30 0.23 0.63 0.37 1.8
Q10 0.19 0.75 0.08 0.11 0.62 0.38 1.2
Q11 0.11 0.81 0.10 0.03 0.67 0.33 1.1
Q12 0.17 0.77 0.05 0.08 0.63 0.37 1.1
Q13 0.09 0.83 0.14 0.06 0.73 0.27 1.1
Q14 0.14 0.74 0.02 0.15 0.59 0.41 1.2
Q15 0.35 0.16 0.74 0.29 0.78 0.22 1.9
Q16 0.31 0.14 0.84 0.14 0.84 0.16 1.4
Q17 0.31 0.02 0.82 0.11 0.78 0.22 1.3
Q18 0.28 0.14 0.78 0.23 0.76 0.24 1.5
Q19 0.23 0.11 0.27 0.77 0.72 0.28 1.5
Q20 0.21 0.14 0.21 0.84 0.81 0.19 1.3
Q21 0.41 0.15 0.12 0.75 0.77 0.23 1.7

                       PC1  PC2  PC3  PC4
SS loadings           5.89 3.42 3.35 2.49
Proportion Var        0.28 0.16 0.16 0.12
Cumulative Var        0.28 0.44 0.60 0.72
Proportion Explained  0.39 0.23 0.22 0.16
Cumulative Proportion 0.39 0.61 0.84 1.00

Mean item complexity =  1.4
Test of the hypothesis that 4 components are sufficient.

The root mean square of the residuals (RMSR) is  0.05

Fit based upon off diagonal values = 0.99
```

*Table 10*

**Computer Code**

```
### Internal consistency
# PC1 (Q1, ..., Q9)
library (psych)
alpha(survey[c(paste("Q",1:9,sep=""))])
```

**Output**

```
Reliability analysis
Call: alpha(x = survey[c(paste("Q", 1:9, sep = ""))])

  raw_alpha std.alpha G6(smc) average_r S/N   ase mean  sd
      0.94      0.94    0.95      0.65  17 0.017  4.5 1.4

 lower alpha upper     95% confidence boundaries
0.91 0.94 0.98
```

```
 Reliability if an item is dropped:
   raw_alpha std.alpha G6(smc) average_r S/N alpha se
Q1      0.93      0.93    0.94      0.64  14    0.020
Q2      0.93      0.93    0.94      0.64  14    0.020
Q3      0.94      0.94    0.94      0.65  15    0.020
Q4      0.94      0.94    0.94      0.66  15    0.020
Q5      0.94      0.94    0.95      0.67  16    0.019
Q6      0.94      0.94    0.94      0.65  15    0.020
Q7      0.94      0.94    0.94      0.65  15    0.020
Q8      0.94      0.94    0.94      0.66  16    0.020
Q9      0.94      0.94    0.94      0.66  15    0.020

 Item statistics
     n raw.r std.r r.cor r.drop mean  sd
Q1 204  0.90  0.90  0.90   0.87  4.4 1.8
Q2 204  0.89  0.89  0.89   0.86  4.5 1.7
Q3 204  0.83  0.84  0.81   0.79  4.9 1.6
Q4 204  0.82  0.81  0.78   0.76  4.6 1.8
Q5 204  0.75  0.76  0.72   0.69  5.1 1.5
Q6 204  0.82  0.83  0.82   0.78  5.0 1.5
Q7 204  0.85  0.84  0.83   0.80  4.5 1.8
Q8 204  0.81  0.80  0.77   0.75  3.6 1.8
Q9 204  0.82  0.81  0.78   0.76  4.2 1.8

Non missing response frequency for each item
      1    2    3    4    5    6    7 miss
Q1 0.08 0.09 0.12 0.19 0.22 0.17 0.13    0
Q2 0.07 0.08 0.11 0.21 0.23 0.17 0.13    0
Q3 0.04 0.04 0.09 0.20 0.22 0.22 0.18    0
Q4 0.07 0.09 0.07 0.22 0.20 0.18 0.17    0
Q5 0.02 0.04 0.06 0.20 0.23 0.28 0.16    0
Q6 0.02 0.03 0.10 0.20 0.22 0.25 0.18    0
Q7 0.09 0.08 0.11 0.15 0.22 0.20 0.15    0
Q8 0.17 0.11 0.20 0.24 0.13 0.07 0.09    0
Q9 0.09 0.11 0.13 0.23 0.20 0.14 0.11    0
```

*Table 11*

*Computer Code*

```
### Internal consistency
# PC2 (Q10...Q14)
library (psych)
alpha(survey[c(paste("Q",10:14,sep=""))])
```

*Output*

```
Reliability analysis
Call: alpha(x = survey[c(paste("Q", 10:14, sep = ""))])

  raw_alpha std.alpha G6(smc) average_r S/N   ase mean   sd
      0.88      0.89    0.88      0.61 7.8 0.035  4.1 0.83

 lower alpha upper     95% confidence boundaries
0.82 0.88 0.95

 Reliability if an item is dropped:
    raw_alpha std.alpha G6(smc) average_r S/N alpha se
Q10      0.87      0.87    0.85      0.63 6.7    0.043
Q11      0.86      0.87    0.84      0.62 6.6    0.043
Q12      0.86      0.86    0.85      0.61 6.2    0.044
Q13      0.85      0.85    0.81      0.58 5.5    0.045
Q14      0.86      0.86    0.83      0.61 6.4    0.044

 Item statistics
      n raw.r std.r r.cor r.drop mean   sd
```

| | |
|---|---|
| | <pre>Q10 204  0.81  0.80  0.73   0.69  4.8 1.08<br>Q11 204  0.82  0.81  0.76   0.70  4.0 1.06<br>Q12 204  0.83  0.83  0.77   0.73  3.9 1.01<br>Q13 204  0.87  0.87  0.85   0.79  4.0 0.92<br>Q14 204  0.81  0.82  0.77   0.71  4.0 0.91<br><br>Non missing response frequency for each item<br>       1    2    3    4    5    6    7 miss<br>Q10 0.05 0.01 0.02 0.05 0.79 0.03 0.04    0<br>Q11 0.05 0.01 0.02 0.81 0.02 0.02 0.05    0<br>Q12 0.06 0.00 0.04 0.80 0.03 0.03 0.02    0<br>Q13 0.04 0.00 0.04 0.82 0.03 0.02 0.03    0<br>Q14 0.04 0.00 0.05 0.83 0.02 0.02 0.03    0</pre> |

*Table 12*

| | |
|---|---|
| *Computer Code* | <pre>### Internal consistency<br># PC2 (Q15...Q18)<br>library (psych)<br>alpha(survey[c(paste("Q",15:18,sep=""))])</pre> |
| *Output* | <pre>Reliability analysis<br>Call: alpha(x = survey[c(paste("Q", 15:18, sep = ""))])<br><br>  raw_alpha std.alpha G6(smc) average_r S/N   ase mean  sd<br>      0.91      0.91    0.89      0.72  10 0.039  4.3 1.6<br><br> lower alpha upper     95% confidence boundaries<br>0.83 0.91 0.99<br><br> Reliability if an item is dropped:<br>    raw_alpha std.alpha G6(smc) average_r S/N alpha se<br>Q15      0.89      0.89    0.85      0.72 7.7    0.054<br>Q16      0.86      0.86    0.81      0.67 6.1    0.056<br>Q17      0.90      0.90    0.86      0.75 8.9    0.052<br>Q18      0.89      0.89    0.84      0.72 7.9    0.053<br><br> Item statistics<br>      n raw.r std.r r.cor r.drop mean  sd<br>Q15 204  0.88  0.88  0.82   0.79  4.6 1.8<br>Q16 204  0.93  0.92  0.91   0.86  4.3 1.8<br>Q17 204  0.86  0.86  0.79   0.75  3.9 1.8<br>Q18 204  0.88  0.88  0.83   0.78  4.5 1.8<br><br>Non missing response frequency for each item<br>       1    2    3    4    5    6    7 miss<br>Q15 0.07 0.08 0.12 0.19 0.19 0.16 0.19    0<br>Q16 0.11 0.06 0.13 0.21 0.16 0.19 0.13    0<br>Q17 0.13 0.09 0.19 0.22 0.16 0.11 0.10    0<br>Q18 0.09 0.07 0.11 0.22 0.18 0.19 0.14    0</pre> |

*Table 13*

| | |
|---|---|
| *Computer Code* | <pre>### Internal consistency<br># PC2 (Q19...Q21)<br>library (psych)<br>alpha(survey[c(paste("Q",19:21,sep=""))])</pre> |
| *Output* | <pre>Reliability analysis<br>Call: alpha(x = survey[c(paste("Q", 19:21, sep = ""))])<br><br>  raw_alpha std.alpha G6(smc) average_r S/N   ase mean  sd<br>      0.83      0.84    0.79      0.64 5.4 0.059    5 1.4<br><br> lower alpha upper     95% confidence boundaries<br>0.72 0.83 0.95<br><br> Reliability if an item is dropped:<br>    raw_alpha std.alpha G6(smc) average_r S/N alpha se<br>Q19      0.82      0.82    0.70      0.70 4.6    0.092<br>Q20      0.70      0.73    0.58      0.58 2.8    0.103<br>Q21      0.76      0.79    0.65      0.65 3.8    0.098<br><br> Item statistics<br>      n raw.r std.r r.cor r.drop mean  sd<br>Q19 204  0.81  0.85  0.73   0.67  5.6 1.2<br>Q20 204  0.91  0.90  0.83   0.76  4.7 1.7<br>Q21 204  0.89  0.87  0.77   0.71  4.5 1.8<br><br>Non missing response frequency for each item<br>       1    2    3    4    5    6    7 miss<br>Q19 0.01 0.01 0.02 0.12 0.21 0.39 0.24    0<br>Q20 0.05 0.07 0.09 0.21 0.21 0.19 0.18    0<br>Q21 0.07 0.09 0.09 0.22 0.23 0.12 0.18    0</pre> |

*Table 14*

| | |
|---|---|
| *Computer Code* | <pre>## CFA with Lavaan package<br>library(lavaan)<br><br>Q.model <-'FACTOR1 =~ Q1 + Q2 + Q3 + Q4 + Q5 + Q6 + Q7 + Q8 + Q9<br>FACTOR2 =~ Q10 + Q11 + Q12 + Q13 + Q14<br>FACTOR3 =~ Q15 + Q16 + Q17 + Q18<br>FACTOR4 =~ Q19 + Q20 + Q21'<br><br><br>fit <- sem(Q.model, data = data.df, std.lv=TRUE,<br>           missing="fiml")<br>summary(fit, fit.measures=TRUE)</pre> |
| *Output* | <pre>lavaan 0.6-2 ended normally after 29 iterations<br><br>  Optimization method                           NLMINB<br>  Number of free parameters                         69<br><br>  Number of observations                           204<br>  Number of missing patterns                         1<br><br>  Estimator                                         ML<br>  Model Fit Test Statistic                     593.222<br>  Degrees of freedom                               183<br>  P-value (Chi-square)                           0.000<br><br>Model test baseline model:<br><br>  Minimum Function Test Statistic             3592.457<br>  Degrees of freedom                               210<br>  P-value                                        0.000<br><br>User model versus baseline model:</pre> |

```
  Comparative Fit Index (CFI)                    0.879
  Tucker-Lewis Index (TLI)                       0.861

Loglikelihood and Information Criteria:

  Loglikelihood user model (H0)              -6269.722
  Loglikelihood unrestricted model (H1)      -5973.111

  Number of free parameters                         69
  Akaike (AIC)                               12677.444
  Bayesian (BIC)                             12906.394
  Sample-size adjusted Bayesian (BIC)        12687.782

Root Mean Square Error of Approximation:

  RMSEA                                          0.105
  90 Percent Confidence Interval          0.095  0.114
  P-value RMSEA <= 0.05                          0.000

Standardized Root Mean Square Residual:

  SRMR                                           0.057

Parameter Estimates:

  Information                                 Observed
  Observed information based on                Hessian
  Standard Errors                             Standard

Latent Variables:
                   Estimate  Std.Err  z-value  P(>|z|)
  FACTOR1 =~
    Q1                1.605    0.096   16.631    0.000
    Q2                1.530    0.094   16.284    0.000
    Q3                1.299    0.094   13.766    0.000
    Q4                1.365    0.105   13.014    0.000
    Q5                1.046    0.091   11.520    0.000
    Q6                1.181    0.088   13.413    0.000
    Q7                1.513    0.105   14.372    0.000
    Q8                1.397    0.106   13.197    0.000
    Q9                1.374    0.103   13.281    0.000
  FACTOR2 =~
    Q10               0.784    0.068   11.443    0.000
    Q11               0.804    0.065   12.286    0.000
    Q12               0.793    0.062   12.869    0.000
    Q13               0.785    0.054   14.652    0.000
    Q14               0.708    0.056   12.633    0.000
  FACTOR3 =~
    Q15               1.520    0.105   14.449    0.000
    Q16               1.676    0.101   16.591    0.000
    Q17               1.417    0.107   13.194    0.000
    Q18               1.517    0.103   14.776    0.000
  FACTOR4 =~
    Q19               0.887    0.075   11.784    0.000
    Q20               1.444    0.104   13.863    0.000
    Q21               1.458    0.109   13.324    0.000

Covariances:
                   Estimate  Std.Err  z-value  P(>|z|)
  FACTOR1 ~~
    FACTOR2           0.378    0.066    5.700    0.000
    FACTOR3           0.676    0.043   15.540    0.000
```

```
    FACTOR4           0.629    0.052   12.098    0.000
  FACTOR2 ~~
    FACTOR3           0.279    0.072    3.877    0.000
    FACTOR4           0.365    0.071    5.114    0.000
  FACTOR3 ~~
    FACTOR4           0.584    0.056   10.361    0.000

Intercepts:
                   Estimate  Std.Err  z-value  P(>|z|)
   .Q1                4.392    0.124   35.400    0.000
   .Q2                4.471    0.120   37.331    0.000
   .Q3                4.892    0.113   43.284    0.000
   .Q4                4.608    0.123   37.373    0.000
   .Q5                5.054    0.103   49.017    0.000
   .Q6                5.010    0.104   47.966    0.000
   .Q7                4.505    0.128   35.216    0.000
   .Q8                3.603    0.125   28.810    0.000
   .Q9                4.176    0.122   34.117    0.000
   .Q10               4.789    0.076   63.338    0.000
   .Q11               4.015    0.074   54.364    0.000
   .Q12               3.936    0.071   55.655    0.000
   .Q13               3.975    0.064   62.041    0.000
   .Q14               3.951    0.064   61.907    0.000
   .Q15               4.578    0.127   36.144    0.000
   .Q16               4.328    0.128   33.689    0.000
   .Q17               3.941    0.126   31.360    0.000
   .Q18               4.461    0.125   35.674    0.000
   .Q19               5.627    0.083   67.920    0.000
   .Q20               4.721    0.120   39.345    0.000
   .Q21               4.529    0.124   36.444    0.000
    FACTOR1           0.000
    FACTOR2           0.000
    FACTOR3           0.000
    FACTOR4           0.000

Variances:
                   Estimate  Std.Err  z-value  P(>|z|)
   .Q1                0.565    0.074    7.650    0.000
   .Q2                0.586    0.073    7.988    0.000
   .Q3                0.917    0.101    9.123    0.000
   .Q4                1.239    0.133    9.315    0.000
   .Q5                1.074    0.112    9.556    0.000
   .Q6                0.831    0.091    9.180    0.000
   .Q7                1.049    0.117    8.932    0.000
   .Q8                1.239    0.133    9.302    0.000
   .Q9                1.170    0.126    9.255    0.000
   .Q10               0.552    0.065    8.551    0.000
   .Q11               0.466    0.057    8.226    0.000
   .Q12               0.391    0.049    8.056    0.000
   .Q13               0.221    0.034    6.454    0.000
   .Q14               0.330    0.041    8.122    0.000
   .Q15               0.964    0.124    7.773    0.000
   .Q16               0.559    0.098    5.718    0.000
   .Q17               1.215    0.140    8.656    0.000
   .Q18               0.889    0.113    7.874    0.000
   .Q19               0.613    0.077    7.925    0.000
   .Q20               0.851    0.142    5.996    0.000
   .Q21               1.026    0.158    6.479    0.000
    FACTOR1           1.000
    FACTOR2           1.000
    FACTOR3           1.000
    FACTOR4           1.000
```

*Table 15*

| | |
|---|---|
| *Computer Code* | `#Adjusted Model`<br>`library(lavaan)`<br><br>`Q.model.2 <-'FACTOR1 =~ Q1 + Q2 + Q3 + Q4 + Q6 + Q7 + Q8 + Q9`<br>`FACTOR2 =~ Q12 + Q13 + Q14`<br>`FACTOR3 =~ Q15 + Q16 + Q17 + Q18`<br>`FACTOR4 =~ Q20 + Q21'`<br><br><br>`fit <- sem(Q.model.2, data = data.df)`<br>`summary(fit, fit.measures=TRUE)` |
| *Output* | see below |

```
lavaan 0.6-2 ended normally after 47 iterations

  Optimization method                           NLMINB
  Number of free parameters                         40

  Number of observations                           204

  Estimator                                         ML
  Model Fit Test Statistic                     386.521
  Degrees of freedom                               113
  P-value (Chi-square)                           0.000

Model test baseline model:

  Minimum Function Test Statistic             2866.310
  Degrees of freedom                               136
  P-value                                        0.000

User model versus baseline model:

  Comparative Fit Index (CFI)                    0.900
  Tucker-Lewis Index (TLI)                       0.879

Loglikelihood and Information Criteria:

  Loglikelihood user model (H0)              -5231.708
  Loglikelihood unrestricted model (H1)      -5038.448

  Number of free parameters                         40
  Akaike (AIC)                               10543.417
  Bayesian (BIC)                             10676.142
  Sample-size adjusted Bayesian (BIC)        10549.410

Root Mean Square Error of Approximation:

  RMSEA                                          0.109
  90 Percent Confidence Interval          0.097  0.121
  P-value RMSEA <= 0.05                          0.000

Standardized Root Mean Square Residual:

  SRMR                                           0.053

Parameter Estimates:

  Information                                 Expected
  Information saturated (h1) model          Structured
  Standard Errors                             Standard
```

```
Latent Variables:
                   Estimate  Std.Err  z-value  P(>|z|)
  FACTOR1 =~
    Q1                1.000
    Q2                0.951    0.047   20.356    0.000
    Q3                0.799    0.051   15.656    0.000
    Q4                0.846    0.057   14.791    0.000
    Q6                0.717    0.048   14.780    0.000
    Q7                0.943    0.055   17.058    0.000
    Q8                0.874    0.057   15.311    0.000
    Q9                0.857    0.056   15.349    0.000
  FACTOR2 =~
    Q12               1.000
    Q13               0.975    0.085   11.525    0.000
    Q14               0.949    0.083   11.408    0.000
  FACTOR3 =~
    Q15               1.000
    Q16               1.108    0.067   16.664    0.000
    Q17               0.934    0.070   13.309    0.000
    Q18               1.001    0.067   14.914    0.000
  FACTOR4 =~
    Q20               1.000
    Q21               1.185    0.119    9.966    0.000

Covariances:
                   Estimate  Std.Err  z-value  P(>|z|)
  FACTOR1 ~~
    FACTOR2           0.435    0.107    4.073    0.000
    FACTOR3           1.648    0.233    7.082    0.000
    FACTOR4           1.376    0.221    6.215    0.000
  FACTOR2 ~~
    FACTOR3           0.307    0.099    3.101    0.002
    FACTOR4           0.383    0.097    3.938    0.000
  FACTOR3 ~~
    FACTOR4           1.097    0.201    5.465    0.000

Variances:
                   Estimate  Std.Err  z-value  P(>|z|)
   .Q1                0.542    0.071    7.676    0.000
   .Q2                0.574    0.072    8.028    0.000
   .Q3                0.947    0.103    9.221    0.000
   .Q4                1.240    0.133    9.348    0.000
   .Q6                0.891    0.095    9.350    0.000
   .Q7                1.028    0.115    8.971    0.000
   .Q8                1.205    0.130    9.274    0.000
   .Q9                1.150    0.124    9.268    0.000
   .Q12               0.399    0.054    7.335    0.000
   .Q13               0.248    0.043    5.796    0.000
   .Q14               0.271    0.043    6.332    0.000
   .Q15               0.977    0.121    8.103    0.000
   .Q16               0.546    0.095    5.756    0.000
   .Q17               1.217    0.139    8.727    0.000
   .Q18               0.888    0.113    7.883    0.000
   .Q20               1.145    0.184    6.236    0.000
   .Q21               0.635    0.212    2.992    0.003
    FACTOR1           2.598    0.310    8.390    0.000
    FACTOR2           0.621    0.100    6.199    0.000
    FACTOR3           2.296    0.318    7.231    0.000
    FACTOR4           1.792    0.304    5.891    0.000
```

*Table 16*

| | |
|---|---|
| *Computer Code* | F1 -> Q1, lam1, NA<br>F1 -> Q2, lam2, NA<br>F1 -> Q3, lam3, NA<br>F1 -> Q4, lam4, NA<br>F1 -> Q5, lam5, NA<br>F1 -> Q6, lam6, NA<br>F1 -> Q7, lam7, NA<br>F1 -> Q8, lam8, NA<br>F1 -> Q9, lam9, NA<br>F2 -> Q10, lam10, NA<br>F2 -> Q11, lam11, NA<br>F2 -> Q12, lam12, NA<br>F2 -> Q13, lam13, NA<br>F2 -> Q14, lam14, NA<br>F3 -> Q15, lam15, NA<br>F3 -> Q16, lam16, NA<br>F3 -> Q17, lam17, NA<br>F3 -> Q18, lam18, NA<br>F4 -> Q19, lam19, NA<br>F4 -> Q20, lam20, NA<br>F4 -> Q21, lam21, NA<br>Q1 <-> Q1, e1, NA<br>Q2 <-> Q2, e2, NA<br>Q3 <-> Q3, e3, NA<br>Q4 <-> Q4, e4, NA<br>Q5 <-> Q5, e5, NA<br>Q6 <-> Q6, e6, NA<br>Q7 <-> Q7, e7, NA<br>Q8 <-> Q8, e8, NA<br>Q9 <-> Q9, e9, NA<br>Q10 <-> Q10, e10, NA<br>Q11 <-> Q11, e11, NA<br>Q12 <-> Q12, e12, NA<br>Q13 <-> Q13, e13, NA<br>Q14 <-> Q14, e14, NA<br>Q15 <-> Q15, e15, NA<br>Q16 <-> Q16, e16, NA<br>Q17 <-> Q17, e17, NA<br>Q18 <-> Q18, e18, NA<br>Q19 <-> Q19, e19, NA<br>Q20 <-> Q20, e20, NA<br>Q21 <-> Q21, e21, NA<br>F1 <-> F1, NA, 1<br>F2 <-> F2, NA, 1<br>F3 <-> F3, NA, 1<br>F4 <-> F4, NA, 1<br>F1 -> F2, F1F2DIR, NA<br>F1 -> F3, F1F3DIR, NA<br>F1 -> F4, F1F4DIR, NA<br>F2 -> F3, F2F3DIR, NA<br>F2 -> F4, F2F4DIR, NA<br>F3 -> F4, F3F4DIR, NA |

*Table 17*

| | |
|---|---|
| *Computer Code* | #with sem<br>library(sem)<br>cfa.model.2<-specifyModel("adjustedmodel.txt")<br>cfaOut.2<-sem(cfa.model.2,S=dataCov,N=204)<br>summary(cfaOut.2) |

*Output*

```
Model Chisquare =  590.31368   Df =  183 Pr(>Chisq) = 2.3016422e-44
 AIC =  686.31368
 BIC =  -382.90228

 Normalized Residuals
       Min.     1st Qu.      Median        Mean     3rd Qu.        Max.
-1.76324751 -0.35685432  0.00000837  0.12497110  0.63869184  2.81328769

 R-square for Endogenous Variables
    Q1     Q2     Q3     Q4     Q5     Q6     Q7     Q8     Q9     F2
Q10    Q11    Q12    Q13    Q14     F3    Q15    Q16    Q17
0.8199 0.7997 0.6480 0.6004 0.5048 0.6264 0.6857 0.6115 0.6173 0.1426
0.5267 0.5814 0.6167 0.7366 0.6026 0.4571 0.7056 0.8340 0.6228
   Q18     F4    Q19    Q20    Q21
0.7213 0.4590 0.5620 0.7101 0.6745

 Parameter Estimates
        Estimate    Std Error   z value     Pr(>|z|)
lam1    1.608582711 0.096797094 16.61808882 5.1549930e-62 Q1 <--- F1
lam2    1.533298744 0.094265385 16.26576646 1.7269253e-59 Q2 <--- F1
lam3    1.302664145 0.094715887 13.75338590 4.8600669e-43 Q3 <--- F1
lam4    1.367859225 0.105220976 12.99987199 1.2254824e-38 Q4 <--- F1
lam5    1.048897739 0.091100541 11.51362793 1.1263809e-30 Q5 <--- F1
lam6    1.183556732 0.088259377 13.40998286 5.2851042e-41 Q6 <--- F1
lam7    1.516650456 0.105613336 14.36040675 9.1687448e-47 Q7 <--- F1
lam8    1.400204107 0.106277223 13.17501596 1.2220159e-39 Q8 <--- F1
lam9    1.377112371 0.103805008 13.26633847 3.6288320e-40 Q9 <--- F1
lam10   0.727571362 0.063832319 11.39816584 4.2702074e-30 Q10 <--- F2
lam11   0.746581481 0.061222614 12.19453777 3.3238467e-34 Q11 <--- F2
lam12   0.736412509 0.057946665 12.70845379 5.3073466e-37 Q12 <--- F2
lam13   0.729131564 0.050471794 14.44631768 2.6445096e-47 Q13 <--- F2
lam14   0.656864351 0.052537122 12.50286122 7.2011974e-36 Q14 <--- F2
lam15   1.122573895 0.082504802 13.60616438 3.6805948e-42 Q15 <--- F3
lam16   1.237868212 0.081432478 15.20116105 3.4740855e-52 Q16 <--- F3
lam17   1.046342686 0.083641448 12.50985862 6.5941490e-36 Q17 <--- F3
lam18   1.120414800 0.081118318 13.81210586 2.1544020e-43 Q18 <--- F3
lam19   0.654116109 0.059562463 10.98201910 4.6638156e-28 Q19 <--- F4
lam20   1.064748301 0.085659917 12.42994786 1.7975157e-35 Q20 <--- F4
lam21   1.074937856 0.088508858 12.14497493 6.1001285e-34 Q21 <--- F4
e1      0.568261817 0.072260791  7.86404086 3.7193614e-15 Q1 <--> Q1
e2      0.589012268 0.072390036  8.13664837 4.0637075e-16 Q2 <--> Q2
e3      0.921920938 0.100352137  9.18685908 4.0447915e-20 Q3 <--> Q3
e4      1.245352450 0.133178260  9.35101909 8.6807243e-21 Q4 <--> Q4
e5      1.079156341 0.112592438  9.58462541 9.2793754e-22 Q5 <--> Q5
e6      0.835549656 0.090168081  9.26657913 1.9221171e-20 Q6 <--> Q6
e7      1.054427368 0.116885817  9.02100355 1.8637404e-19 Q7 <--> Q7
e8      1.245527888 0.133693832  9.31627039 1.2050172e-20 Q8 <--> Q8
e9      1.175714719 0.126456396  9.29739224 1.4393243e-20 Q9 <--> Q9
e10     0.554733476 0.063445756  8.74342923 2.2613954e-18 Q10 <--> Q10
e11     0.467959800 0.055688186  8.40321500 4.3441829e-17 Q11 <--> Q11
e12     0.393014520 0.048344009  8.12953929 4.3092507e-16 Q12 <--> Q12
e13     0.221743796 0.033134028  6.69232836 2.1964731e-11 Q13 <--> Q13
e14     0.331820883 0.040244589  8.24510562 1.6501439e-16 Q14 <--> Q14
e15     0.968292549 0.120364329  8.04468035 8.6470830e-16 Q15 <--> Q15
e16     0.561857872 0.095905395  5.85845952 4.6718045e-09 Q16 <--> Q16
e17     1.221343907 0.140347007  8.70231532 3.2518057e-18 Q17 <--> Q17
e18     0.893177527 0.113469641  7.87151100 3.5038332e-15 Q18 <--> Q18
e19     0.616437482 0.076967253  8.00908772 1.1556246e-15 Q19 <--> Q19
e20     0.855556466 0.141520582  6.04545612 1.4898790e-09 Q20 <--> Q20
e21     1.030772948 0.155468876  6.63009200 3.3547771e-11 Q21 <--> Q21
F1F2DIR 0.407736694 0.083275769  4.89622252 9.7696435e-07 F2 <--- F1
F1F3DIR 0.902724046 0.112598958  8.01716163 1.0821668e-15 F3 <--- F1
F1F4DIR 0.520806176 0.131646222  3.95610423 7.6181932e-05 F4 <--- F1
```

| | |
|---|---|
| | F2F3DIR 0.034731466 0.082356936  0.42171878 6.7323029e-01 F3 <--- F2<br>F2F4DIR 0.176535492 0.088746145  1.98921871 4.6677067e-02 F4 <--- F2<br>F3F4DIR 0.286376634 0.093034567  3.07817453 2.0827290e-03 F4 <--- F3<br><br>Iterations =  36 |